\documentclass{emulateapj}
\usepackage{subfigure}
\def\SP{\let\\=\empty\futurelet\C\puncspace }

\def\micron{$\mu$m}

\def\iras{{\it{IRAS}}}
\usepackage{rotating}

\begin{document}

\title{$Spitzer$ Observations of Cold Dust Galaxies}
\author{
  C. N. A. Willmer\altaffilmark{1},
  G. H. Rieke\altaffilmark{1},
  Emeric Le Floc'h\altaffilmark{2,3},
  J. L. Hinz\altaffilmark{1},
  C. W. Engelbracht\altaffilmark{1}
  Delphine Marcillac\altaffilmark{4},
  and K. D. Gordon\altaffilmark{5},
}
\altaffiltext{1}{Steward Observatory, University of Arizona, 933
  North Cherry Avenue, Tucson, AZ 85721 ;cnaw@as.arizona.edu}
\altaffiltext{2}{Spitzer Fellow}
\altaffiltext{2}{Institute for Astronomy, University of Hawai'i
  2680 Woodlawn Drive, Honolulu, HI 96822-1839}
\altaffiltext{4}{Institut d'Astrophysique Spatiale, B\^at. 121, Universit\'e de
  Paris Sud XI, 91405 Orsay Cedex, France}
\altaffiltext{5} {Space Telescope Science Institute, 3700 San Martin Dr.
Baltimore, MD 21218}

\begin{abstract}
We combine new {\it{Spitzer Space Telescope}} observations in the mid- and
far-infrared with SCUBA 850\micron~ observations to improve the
measurement of dust temperatures, masses and luminosities for 11
galaxies of the SCUBA Local Universe Galaxy Survey (SLUGS).
By fitting dust models we measure typical dust masses of
10$^{7.9}M_\odot$ and dust luminosities of $\sim$10$^{10}L_\odot$, for
galaxies with modest star formation rates.
The data presented in this paper combined with
previous observations show that cold dust is present in all types of
spiral galaxies and is a major contributor to their total luminosity.
Because of the lower dust temperature of the SCUBA sources measured in
this paper, they have flatter Far-IR
$\nu$F$_\nu$(160$\mu$m)/$\nu$F$_\nu$(850$\mu$m) slopes than the larger
{\it{Spitzer Nearby Galaxies Survey}} (SINGS), the sample that 
provides the best measurements of the dust properties of galaxies in
the nearby universe. The new data presented here added to
SINGS extend the parameter space that is well 
covered by local galaxies, providing a
comprehensive set of templates that can be used to interpret the
observations of nearby and distant galaxies.

\end{abstract}

\clearpage

\section{Introduction}

The mid and far infrared  ($\sim$ 10-1000 \micron) is a wavelength
domain where some galaxy populations show significant changes over time
(e.g., Puget et al. 1996; Sanders \& Mirabel 1996; Chary \& Elbaz
2001; Elbaz et al. 2002; Le Floc'h et al. 2005; Lagache, Dole \& Puget
2005) and is therefore 
critical in our understanding of how galaxies form and evolve. 
The strong emission of galaxies in
these wavelengths has been measured for objects with very different
optical properties (spirals, ellipticals, Seyferts, interacting
galaxies). The optical to IR ratio varies strongly from galaxy to
galaxy and can be as low as few percent for the extreme IR galaxies
($L_{IR} > 10^{11} L_\odot$) where the IR emission dominates their
total luminosity (Rieke \& Lebofsky 1978). 

This IR emission is mostly due to starlight absorbed by dust grains
and re-emitted at longer wavelengths, though some objects have a
significant non-thermal contribution coming from ionized gas in the interstellar
medium (ISM) (Laurent et al. 2000a) and from active galactic
nuclei and radio sources. The mid-IR light ($\lambda \sim$ 5--30 \micron) 
is emitted by  dust grains of different sizes (D\'esert, Boulanger \& Puget
1990), that include very small grains (VSG), which are responsible
for the thermal emission at 10-16 $\mu$m and are located in HII regions
(Laurent et al. 2000b), and aromatic molecules (e.g., PAH,
Leger \& Puget 1984) that absorb ultra-violet photons generated in
photo-dissociation regions (PDR) and  re-emit in several bands between 3 and 17
\micron~ (e.g., Laurent et al. 2000b; Lagache, Puget \& Dole 2005).
At longer wavelengths the emission is due to larger grains composed
mainly of silicates (D\'esert et al. 1990). The latter absorb light
from the ISM that is generated by young OB
associations and from the old disk population (Lonsdale \&
Helou 1987). The ubiquity of dust in galaxies was demonstrated by the
Infrared Astronomical Satellite (\iras) ~survey,  which covered over
90\% of the sky in wavelengths 
ranging from 12~\micron~ to 100~\micron, and showed that most of the
detected infrared luminosity could be explained by dust at
temperatures of 40K and higher.  
However, evidence for dust at more than one temperature was first
noted by Helou (1986) from the $IRAS$ 12/25 $\mu$m to 60/100 $\mu$m
color-color diagram which suggested that in addition to dust excited
by the actively star
forming regions, another component due to cooler ``cirrus'' was
present and motivated the models proposed by Lonsdale \& Helou (1987)
mentioned above, and validated by Sauvage \& Thuan (1992) in study
comparing the H$\alpha$ and far-IR luminosities. From the
analysis of the gas to dust ratio in spiral galaxies, Devereux \&
Young (1990) could explain the observed discrepancy between the gas to
dust ratios measured in external galaxies with the measurements
obtained for the Galaxy by assuming the presence of a diffuse and cold
(T $<$ 30K) dust component dominant in the outer regions of galaxies,
while the warmer dust would be predominant in the inner regions of the
galaxy.

The detection of this cold dust
required reaching longer wavelengths than accessible with \iras.
This became possible with the flight of the Infrared Space Observatory
(ISO, Kessler et al. 1996) which allowed probing out to 200 \micron~
and with the commissioning of the Submillimeter High Angular
Resolution Camera (SHARC) at the Caltech Submilliter Observatory
(e.g., Serabyn et al. 1999) and on the James Clerk Maxwell Telescope of
the Submillimetre Common-User Bolometer Array (SCUBA) (Holland et 
al. 1999), observing at 850 \micron. 

%
%
\begin{table*}
\scriptsize
\caption{Galaxy sample}
\begin{center}
\begin{tabular}{lcrrrrr}
\hline
\hline
Galaxy     &    z    & T&  M$_B$ & Log$_{10}$( L$_{IR}$)&  Log$_{10}$(L$_{TIR}$)&SFR\cr
 {}        &   {}    &{}& {}     & $L_{\odot} $ & $L_{\odot} $&
 $M_{\odot}$ yr$^{-1}$\cr
\noalign{\smallskip} \hline \noalign{\smallskip}
IC~797     & 0.00320 & 6 & -16.81 &   9.27&   8.81& 0.2 \cr
IC~800     & 0.00779 & 5 & -18.83 &   9.07&   9.38& 0.4\cr
NGC~803    & 0.00703 & 5 & -19.28 &   9.37&   9.63& 4.5\cr
NGC~3270   & 0.02112 & 3 & -21.22 &  10.23&  10.56& 2.0\cr
NGC~4712   & 0.01473 & 4 & -20.91 &   9.87&  10.15& ...\cr
NGC~6137   & 0.03153 &-5 & -21.98 &$<$9.94&$<$9.54& 2.9\cr
NGC~7047   & 0.01957 & 3 & -20.50 &   9.98&  10.35& 3.5\cr
NGC~7653   & 0.01433 & 3 & -20.48 &  10.17&  10.43& 0.7\cr
PGC~035952 & 0.01331 & 4 & -19.73 &   9.61&   9.80& 1.0\cr
UGC~10205  & 0.02211 & 1 & -20.80 &  10.10&  10.35& 2.9\cr
UGC~12519  & 0.01471 & 5 & -19.76 &   9.95&  10.30& 2.7\cr
\noalign{\smallskip} \hline
\end{tabular}
\end{center}
Note. -- Redshifts and morphologies come from the NASA Extragalactic
Database (NED). The absolute magnitudes were derived from B magnitudes
in NED. The  L$_{IR}$ measurements come from Vlahakis et al. (2005),
while the L$_{TIR}$ are derived from the the fits described in Section
3. The star formation rates are estimated following Bell (2003) and
have estimated uncertainties $\sim$ 50\%.
\end{table*}

The ISO observations of the nearby galaxies M~31 (Haas et al. 1998), M~33
(Hippelein et al. 2003) and M~101 (Tuffs \& Gabriel 2003; Sauvage,
Tuffs \& Popescu 2005) and by Serabyn et al. (1999) for NGC~891 using
SHARC, in wavelengths above 100 \micron~ provided the
first direct detections of cold dust. 
In a study of NGC~891, Popescu
\& Tuffs (2003) showed that the cold dust extends beyond the stellar
disk, a similar behavior observed in HI
disks, thus confirming the scenario proposed by Devereux \& Young
(1990). 
The data acquired for M~31  (Haas et al. 1998), M~101 (Tuffs
\& Gabriel 2003), M~33 (Haas et al. 1998; Hinz et al. 2004) and NGC~55
(Engelbracht et al. 2004) confirm that the cold and warm dust are
distributed differently throughout the galaxy. In general, the warm
dust component is easily identified in spiral arms and nuclei thanks
to the larger density of star-forming regions, while the
cold component is the source of the diffuse emission in the disk,
though some localized components exist in the spiral arms (Tuffs \&
Popescu 2005).

The ground based observations using SCUBA to target nearby galaxies with
\iras ~detections also suggested the presence of significant amounts of cold
dust. 
Dunne et al. (2000) obtained SCUBA
observations for 104 galaxies and were able to obtain
measurements of the local submillimeter and dust mass
functions. However, because the original sample of 
Dunne et al. (2000) was selected on the basis of \iras ~flux, they
would preferentially detect galaxies with a significant amount of dust
at T$>$40K. Based on the high gas-to-dust ratio ($\sim$ 600) compared
to the Galactic value, Dunne et al. (2000) suggested that a cold dust
component would be present in their sample, but which would not be
directly detected by $IRAS$. To overcome the bias of using only
$IRAS$-selected galaxies,
Vlahakis, Dunne \& Eales (2005) used the CfA1 redshift survey
(Huchra et al. 1983) to define a sample to observe with SCUBA based
exclusively on their optical properties. These were observed at 850
$\mu$m as part of the 
Scuba Local Universe Galaxy Survey (SLUGS).
The SLUGS sample confirmed previous findings (e.g., Tuffs \& Popescu
2003) using ISO that most of the dust mass is contained in cold
particles. Whenever SCUBA  data at 450 $\mu$m were
available, supplementing the 850 $\mu$m and $IRAS$ measurements, Vlahakis et
al. (2005) found that two dust components (one warm, another cold) generally
produced better fits to the observed distribution of Mid- to Far-IR
fluxes. Their fits suggested that the total cold dust mass in these
galaxies could be up to 100 times larger than the dust responsible for
the emission at 5 $\mu$m through 70 $\mu$m.

Our understanding of dust properties of galaxies in the nearby
universe has increased dramatically with the {\it{Spitzer Nearby
Galaxies Survey}} (SINGS) key
project (Kennicutt et al. 2003). SINGS has obtained 
extensive observations ranging from the far-UV observed with the
Galaxy Evolution Explorer (GALEX, Martin et al. 2005)
to the far-infrared for a representative sample of nearby galaxies
comprising a wide range of luminosities, sizes and morphologies (e.g.,
Dale et al. 2005; Dale et al. 2007; Draine et al. 2007; Smith et
al. 2007). These objects were pre-selected considering a parameter
space defined by the optical morphology, the luminosity and the
optical/FIR luminosity ratio (Kennicutt et al. 2003). Because
prior to the launch of $Spitzer$ most of the available mid- and far-IR
measurements used to define the sample were obtained by \iras, mainly
sensitive to the emission by hot and warm dust, it is possible
that galaxies with large amounts of cold dust are
under-represented in SINGS. To test if this bias is present, we will use
galaxies from SLUGS to address the far-IR properties of galaxies with
a large cold dust component, and to provide a more complete view of
the properties of galaxies over the whole extent of the IR wavelengths.
In Section 2 we present the sample and describe the
observations. In Section 3 we use modified black-body curves and
dust models to estimate the temperatures of the dust components and characterize 
the Spectral Energy Distributions of these galaxies. Section 4 presents 
our conclusions. An $H_0$= 70 $km s^{-1} Mpc^{-1}$, $\Omega_m$ = 0.3
and $\Omega_\Lambda$=0.7 cosmology is assumed throughout.

\section{Data}
%
%
\begin{table*}
\scriptsize
\caption{Detector Parameters}
\begin{center}
\begin{tabular}{lrrrcc}
\hline
\hline
Detector & $\lambda$$_0$~ & Pixel Size  & FWHM    & Conversion Factor
& Calibration Uncertainty\cr
 {}      &   $\mu$m~          & arc sec & arc sec & Jy/(MJy/Sr) &{} \cr
\noalign{\smallskip} \hline \noalign{\smallskip}
IRAC 3.6 & 3.550            & 0.86 & 1.66& 1.7492$\times$10${-5}$ & 1.8\%\cr
IRAC 4.5 & 4.493            & 0.86 & 1.72& 1.7492$\times$10${-5}$ & 1.9\% \cr
IRAC 5.8 & 5.731            & 0.86 & 1.88& 1.7492$\times$10${-5}$ & 2.0\%\cr
IRAC 8.0 & 7.872            & 0.86 & 1.98& 1.7492$\times$10${-5}$ & 2.1\%\cr
MIPS 24  & 23.68            & 2.49 & 6.0     & 6.514$\times$10${-6}$&4\% \cr 
MIPS 70  & 71.42            & 9.85 & 18      & 1.60 &5\%\cr
MIPS 160 & 155.90           & 16   & 40      & 0.25 &15\%\cr
\noalign{\smallskip} \hline
\end{tabular}
\end{center}
\end{table*}

\subsection{Sample definition}

The $Spitzer$ observations targeted a subsample of 11 galaxies from
Vlahakis et al. (2005) that have SCUBA observations in 850 \micron~
(and in three cases additional data at 450 \micron). The sample,
presented in Table 1, contains
galaxies with morphologies ranging from elliptical to late spiral, but
with no cases of irregular or interacting galaxies. All
galaxies in this sample are relatively quiescent in terms of
their star formation activity with $\sim$ 2 $M_{\odot}$ yr$^{-1}$,
estimated using the Bell (2003) relation. Because these galaxies are
at moderately low redshift, with typical sizes of $\sim 2'$, the main subcomponents
(bulge, disk, possibly spiral arms) are easily resolved in data
ranging from the optical down to 24 \micron.

\subsection{Observations}

The observations were taken as part of the
$Spitzer$ Guaranteed Time Observation Program 30348 (P.I.: G. H. Rieke)
and used both the Infrared Array Camera (IRAC, Fazio et
al. 2004), which is sensitive to the Rayleigh-Jeans stellar
and warm dust emission, and the Mid Infrared Photometer for $Spitzer$
(MIPS, Rieke et al. 2004), which is sensitive to warm dust in its
bluer wavelength windows (24 \micron~ and 70 \micron) and cold dust at
160 \micron.

The IRAC data were taken
using all four wavelengths (3.6, 4.5, 5.8 and 8.0 \micron), where a
5$' \times 5'$ region of the sky was observed, using a 5 position
dithered pattern, each exposed for 30 seconds, giving a total
exposure of 150 seconds per band. 

The MIPS observations were taken in photometry mode, where the
spacecraft points at the object and then performs a series of mirror
motions and spacecraft maneuvers to dither the source over the array
(Rieke et al. 2004). All MIPS 24 \micron~ observations were taken in a
single cycle using a sequence of 12 dithers, followed
by another series of 12 images which are offset to calculate the sky
background. The total on-source integration is 120 seconds. Most 70
$\mu$m observations were also taken in a single cycle, though 
for two galaxies, because of their large apparent extent, ``fixed cluster''
observations using two cycles were necessary; the total integration
time per pixel amounts to $\sim$ 125 seconds. In the case of 160
$\mu$m, the observations were taken over 4 cycles, which adds to a
total integration time of $\sim$ 42 seconds per pixel.

Table 2 summarizes the instrumental parameters for each detector,
consisting of the nominal wavelength (e.g., Reach et al. 2005; Rieke
et al. 2008), the pixel size in the final mosaic, the FWHM of the
point-spread function, the factor used to convert fluxes into
Janskys and the calibration uncertainty.

\begin{figure*}
\vspace{22cm}
\includegraphics{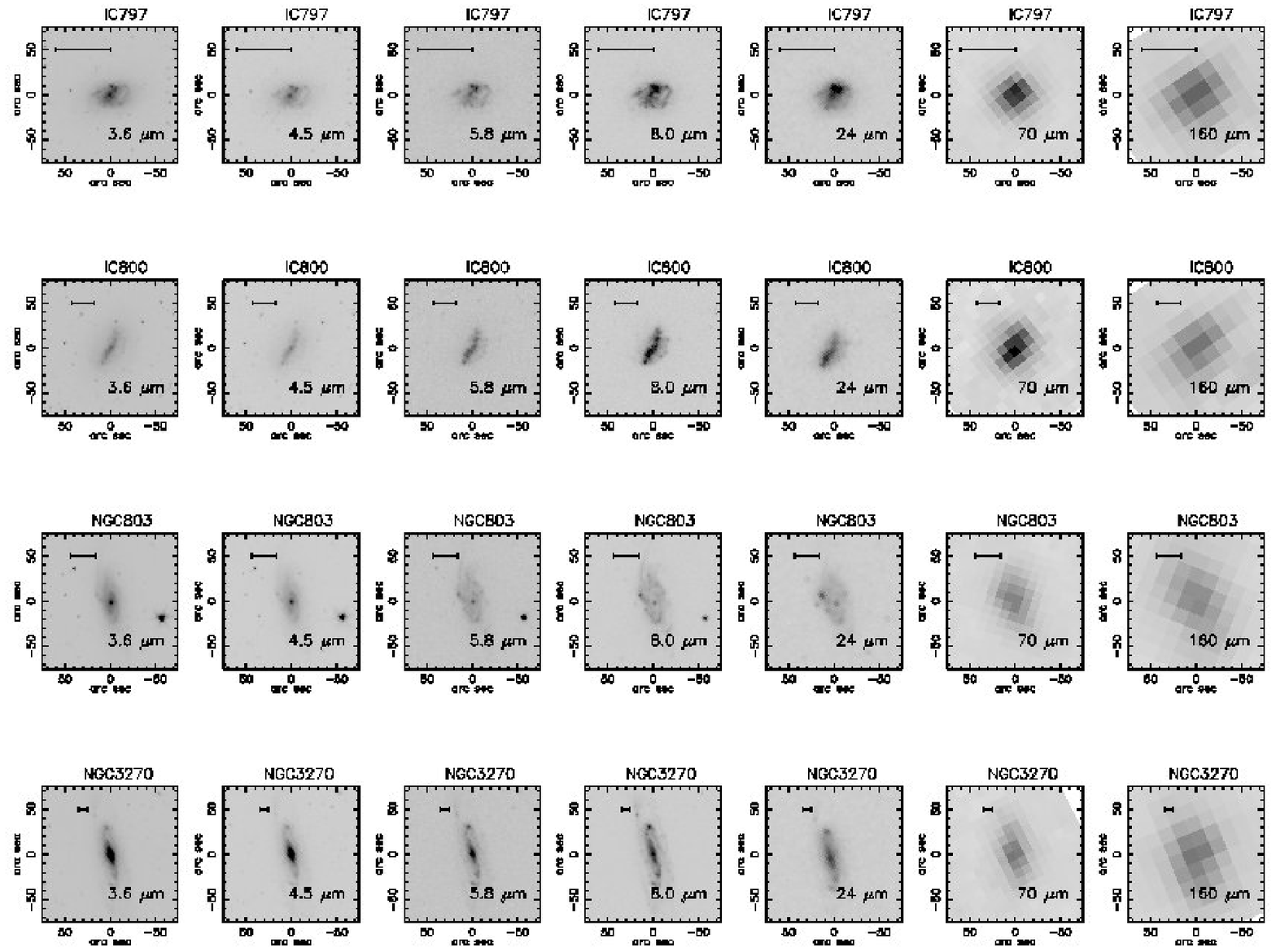}
\includegraphics{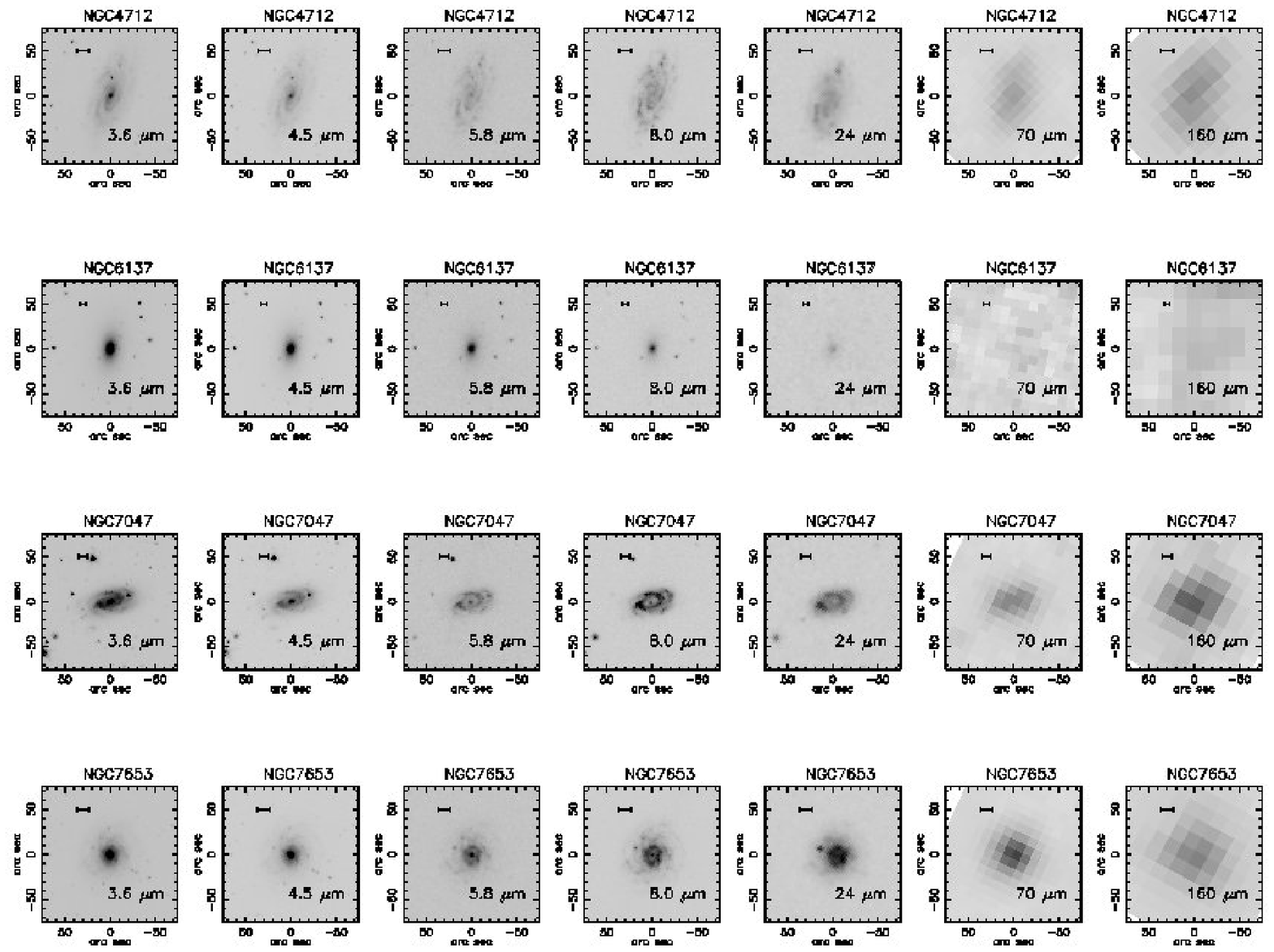}
\caption{Postage stamp images of galaxies in this sample. The galaxy
  identification is noted at top of window, while the band is noted in
  the lower right corner. A scale bar corresponding to 2 kpc at the
  galaxy distance is shown in the upper left corner of each figure.
}
\end{figure*}

\subsubsection{Data Reduction}

The pipeline-processed IRAC Basic Calibrated Data (BCD) were retrieved
from the {\it{Spitzer Science Center}} (SSC) and were post-processed
using the IRACproc suite of scripts 
(Schuster et al. 2006). These combine the SSC-processed images,
reject outliers, correct for bad-pixels and cosmic rays, and 
produce the final flux-calibrated mosaic (Schuster et al. 2006), using
the procedure and calibrations of Reach et al. (2005). The final
photometry includes the calibration corrections determined by Rieke et
al. (2008). The photometric 
uncertainties in the IRAC calibration are  1.8\%, 1.9\%, 2.0\%
and 2.1\% for  3.6 $\mu$m, 4.5 $\mu$m, 5.8 $\mu$m and 8.0 $\mu$m
respectively (Reach et al. 2005). The IRAC measurements already
contain the corrections for extended
sources\footnote{\url{http://ssc.spitzer.caltech.edu/irac/calib/extcal/index.html}}.
No color corrections were applied to
the tabulated data.

The MIPS data were reduced using the Data Analysis Tool version 3.06
(DAT) which is extensively described by Gordon et al. (2005). The DAT 
reads the raw data downloaded from the SSC
archive, converts the integration ramps into slopes (in data numbers
per second) removing at the same time instrumental effects, electronic
non-linearities and stimulator flash latents in the case of 70 and
160 $\mu$m observations. The DAT also performs dark current,
flat field and illumination corrections, removes distortions
introduced by the off-axis location of the detectors, and creates a
flux-calibrated mosaic (Gordon et al. 2005). The calibration of the
MIPS data, described by Engelbracht et al. (2007), Gordon et
al. (2007) and Stansberry et al. (2007), used repeated observations
of stars which are referred to a model spectral energy distribution of
an ideal A star (``Vega''), calibrated using the values in Rieke et al. (2008). The
calibration uncertainties are $\sim$ 4\% for 24 $\mu$m (Engelbracht
et al. 2007), 5 \% for 70 $\mu$m (Gordon et al. 2007) and $\sim$ 15
\% for 160 $\mu$m (Stansberry et al. 2007).

\addtocounter{figure}{-1} 
\begin{figure*}
\vspace{8cm}
\includegraphics{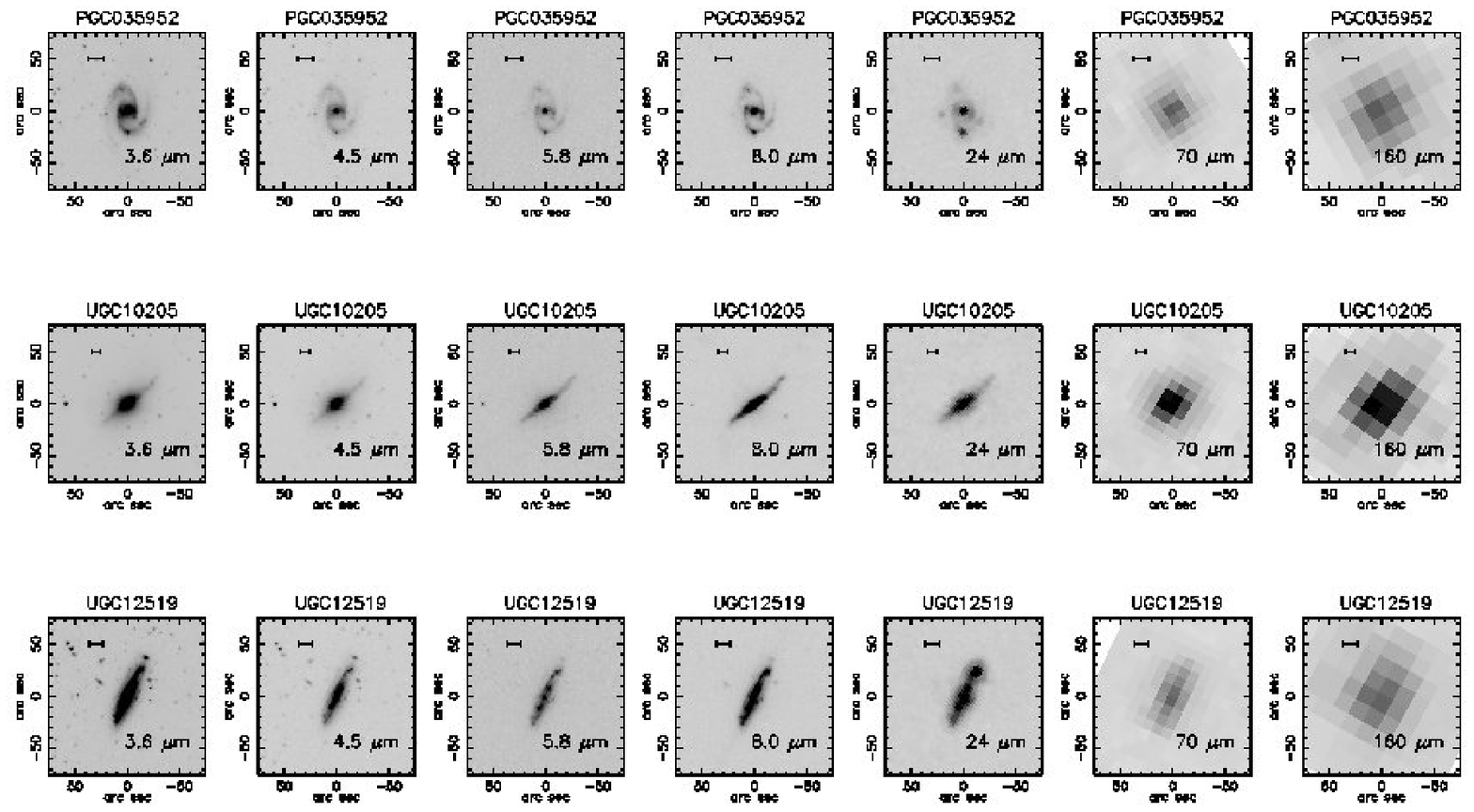}
\caption{ (cont.)
}
\end{figure*}
\begin{table*}
\scriptsize
\caption{Galaxy sample: $Spitzer$ Measurements}
\begin{center}
\begin{tabular}{lcccccccc}
\hline
\hline
Name & 3.6$\mu$m & 4.5$\mu$m & 5.8$\mu$m & 8.0$\mu$m & 24$\mu$m &
70$\mu$m & 160$\mu$m \cr
 {}  & Jy & Jy & Jy & Jy & Jy & Jy & Jy\cr
\hline
IC~797 &    0.021$\pm$   0.003 &    0.014$\pm$   0.002 &    0.027$\pm$   0.004 &    0.065$\pm$   0.010 &    0.072$\pm$   0.007 &    0.939$\pm$   0.188 &    2.708$\pm$   0.677\cr
IC~800 &    0.017$\pm$   0.003 &    0.012$\pm$   0.002 &    0.020$\pm$   0.003 &    0.042$\pm$   0.006 &    0.046$\pm$   0.005 &    0.519$\pm$   0.104 &    1.463$\pm$   0.366\cr
NGC~803 &    0.045$\pm$   0.007 &    0.029$\pm$   0.004 &    0.042$\pm$   0.006 &    0.092$\pm$   0.014 &    0.071$\pm$   0.007 &    0.904$\pm$   0.181 &    4.241$\pm$   1.060\cr
NGC~3270 &    0.050$\pm$   0.007 &    0.031$\pm$   0.005 &    0.043$\pm$   0.006 &    0.103$\pm$   0.015 &    0.083$\pm$   0.008 &    0.725$\pm$   0.145 &    3.727$\pm$   0.932\cr
NGC~4712 &    0.028$\pm$   0.004 &    0.019$\pm$   0.003 &    0.027$\pm$   0.004 &    0.071$\pm$   0.011 &    0.069$\pm$   0.007 &    0.711$\pm$   0.142 &    3.236$\pm$   0.809\cr
NGC~6137 &    0.058$\pm$   0.009 &    0.036$\pm$   0.005 &    0.022$\pm$   0.003 &    0.018$\pm$   0.003 &    0.007$\pm$   0.001 &    0.017$\pm$   0.003 &    0.030$\pm$   0.007\cr
NGC~7047 &    0.028$\pm$   0.004 &    0.019$\pm$   0.003 &    0.031$\pm$   0.005 &    0.067$\pm$   0.010 &    0.060$\pm$   0.006 &    0.722$\pm$   0.144 &    2.201$\pm$   0.550\cr
NGC~7653 &    0.046$\pm$   0.007 &    0.030$\pm$   0.005 &    0.060$\pm$   0.009 &    0.145$\pm$   0.022 &    0.131$\pm$   0.013 &    1.776$\pm$   0.355 &    4.346$\pm$   1.087\cr
PGC~035952 &    0.011$\pm$   0.002 &    0.006$\pm$   0.001 &    0.011$\pm$   0.002 &    0.028$\pm$   0.004 &    0.045$\pm$   0.005 &    0.611$\pm$   0.122 &    1.211$\pm$   0.303\cr
UGC~10205 &    0.042$\pm$   0.006 &    0.028$\pm$   0.004 &    0.029$\pm$   0.004 &    0.056$\pm$   0.008 &    0.041$\pm$   0.004 &    0.727$\pm$   0.145 &    1.762$\pm$   0.441\cr
UGC~12519 &    0.020$\pm$   0.003 &    0.013$\pm$   0.002 &    0.033$\pm$   0.005 &    0.078$\pm$   0.012 &    0.108$\pm$   0.011 &    1.081$\pm$   0.216 &    3.690$\pm$   0.922\cr
\hline
\end{tabular}
\end{center}
\end{table*}

\begin{table*}
\scriptsize
\caption{Galaxy sample: Measures from the Literature}
\begin{center}
\begin{tabular}{lrrrccccc}
\hline
\hline
Name &  J~~~~~~ & H~~~~~~ & K~~~~~~ & 12$\mu$m & 60$\mu$m & 100$\mu$m & 450$\mu$m & 850$\mu$m\cr
 {}  & mag~~~~ & mag~~~~~ & mag~~~~ & Jy & Jy & Jy & Jy & Jy\cr
\hline
IC~797     & 11.58$\pm$0.04 & 11.07$\pm$0.06 & 10.80$\pm$0.07 &    0.200$\pm$   0.043 &    0.730$\pm$   0.045 &    2.110$\pm$   0.130 & ... & 0.085$\pm$0.021\cr
IC~800     & 11.19$\pm$0.03 & 11.34$\pm$0.03 & 10.58$\pm$0.06 &    0.130$\pm$   0.038 &    0.600$\pm$   0.056 &    1.560$\pm$   0.271 & ... & 0.076$\pm$0.019\cr
NGC~803    & 10.71$\pm$0.02 & 10.01$\pm$0.03 & 9.83$\pm$0.04 &    0.140$\pm$   0.040 &    0.730$\pm$   0.047 &    2.850$\pm$   0.134 & 0.631$\pm$0.22085 & 0.093$\pm$0.019\cr
NGC~3270   & 10.76$\pm$0.02 & 10.06$\pm$0.01 & 9.73$\pm$0.02 &    0.130$\pm$   0.032 &    0.620$\pm$   0.040 &    2.280$\pm$   0.107 & ... & 0.059$\pm$0.014\cr
NGC~4712   & 10.92$\pm$0.04 & 10.42$\pm$0.05 & 10.09$\pm$0.05 &    0.070$\pm$   0.033 &    0.500$\pm$   0.044 &    1.880$\pm$   0.113 & ... & 0.102$\pm$0.023\cr
NGC~6137   & 10.51$\pm$0.02 & 9.81$\pm$0.02 & 9.55$\pm$0.03 &    0.020$\pm$   0.020 &    0.100$\pm$   0.100 &    0.380$\pm$   0.380 & ... & 0.029$\pm$0.010\cr
NGC~7047   & 11.13$\pm$0.03 & 10.47$\pm$0.04 & 10.20$\pm$0.05 &    0.120$\pm$   0.025 &    0.580$\pm$   0.059 &    2.110$\pm$   0.251 & ... & 0.055$\pm$0.013\cr
NGC~7653   & 10.71$\pm$0.02 & 10.03$\pm$0.02 & 9.75$\pm$0.04 &    0.280$\pm$   0.041 &    1.440$\pm$   0.065 &    5.460$\pm$   0.434 & ... & 0.112$\pm$0.020\cr
PGC~035952 & 12.26$\pm$0.05 & 11.58$\pm$0.07 & 11.29$\pm$0.07 &    0.030$\pm$   0.034 &    0.420$\pm$   0.047 &    1.600$\pm$   0.212 & 0.421$\pm$0.14735 & 0.051$\pm$0.013\cr
UGC~10205  & 10.87$\pm$0.02 & 10.14$\pm$0.02 & 9.89$\pm$0.03 &    0.060$\pm$   0.041 &    0.460$\pm$   0.056 &    1.560$\pm$   0.192 & ... & 0.058$\pm$0.015\cr
UGC~12519  & 11.62$\pm$0.03 & 10.89$\pm$0.03 & 10.61$\pm$0.06 &    0.130$\pm$   0.034 &    0.960$\pm$   0.052 &    3.280$\pm$   0.321 & 0.408$\pm$0.1428 & 0.074$\pm$0.016\cr
\hline
\end{tabular}
\end{center}
Note. -- J, H and K magnitudes come from the 2MASS archive at
      {\url{http://irsa.ipac.caltech.edu/}}; the \iras ~measurements
      were obtained using the Scanpi tool at 
      {\url{http://scanpiops.ipac.caltech.edu:9000/applications/Scanpi/index.html}}.
      The 450 and 850 $\mu$m measurements come from Vlahakis et al. (2005)\\
\end{table*}
\subsubsection{Photometry}

Galaxy fluxes were measured using the IRAF{\footnote{IRAF is
  distributed by the National Optical Astronomy Observatory, which is
  operated by the Association of Universities for Research in
  Astronomy (AURA), Inc., under cooperative agreement with the
  National Science Foundation.}} $ellipse$ task (Busko
1996), based on the method of Jedrzejewski (1987) to fit 
elliptical and circular isophotes. This task takes first guess values of
object center, semi-major axis, position angle and ellipticity to
calculate the ellipse at the starting radius. Excepting for the
semi-major axis, the fitted parameters are used as input values for
the next radius (outer or inner), which is calculated using a radial step
set by the user. 
The $ellipse$ task allows masking bad pixels as well as other objects in the
image, such as foreground stars and background galaxies.
Since $ellipse$ does not calculate sky background values, these must be
subtracted prior to fitting the ellipses.

For the IRAC data, the initial parameters were obtained from the 
Bertin \& Arnouts (1995) source extractor (SExtractor) catalog
derived for each image, while the segmentation image produced by
SExtractor was used to mask out stars and background galaxies. To
produce a valid ``data quality file'' input to $ellipse$ all pixels
belonging to the object and the sky were set to zero in this image. 
In addition to masking foreground and background sources, the ``data
quality file''  is also used to flag pixels affected by
instrumental defects and cosmic rays.
The same procedure to mask background/foreground objects was used
for the MIPS observations. In the case of 
MIPS 160 $\mu$m images, gaps in the sky that were not covered by the
array were masked out by hand. 

Prior to the photometry, fluxes were converted from MJy/Sr into
Janskys using the factors presented in Table 2, which depend on the
pixel size.
The sky level in the IRAC images was determined by fitting a low 
order polynomial in regions more than 100$''$ away from the galaxies,
which was then subtracted from the image. For MIPS images the
background sky was estimated using several boxes of 25 pixels from
which a mean sky value was determined and subtracted. For most
galaxies the surface brightness profile flattens at large radii,
demonstrating that the sky subtraction was effective. However, in a few
cases, in particular for 160 $\mu$m, the presence of residual sky
(caused by the foreground Galactic Cirrus) can
affect the final measurements of galaxies with very low flux in this
band.  The  circular and elliptical isophotes were fit to distances of $\sim$
100 $''$ from the center and for most bands convergence was reached at 
about 50$''$. For most of the sources considered
in this work, the results derived from elliptical and  circular
apertures differ by less than the estimated uncertainties. Based on this,
the circular aperture photometry was adopted for the final flux
measurements, which are presented in Table 3.

\subsection{Archival data of other wavelengths}

The $Spitzer$ IRAC and MIPS data are supplemented in this paper with
published data, presented in Table 4, consisting of the 2MASS
photometry in the near- infrared, the sub-mm measurements
by Vlahakis et al. (2005) and \iras ~archival data.

The 2MASS (Jarrett et al. 2000) near-Infrared magnitudes were obtained
from the NASA Extragalactic Database (NED). The 2MASS measurements are
the total magnitudes measured by extrapolating 
the near-IR surface brightness to a diameter corresponding
to roughly 4 disk scale lengths defined from the $J$-band image due to
its greater sensitivity at lower surface brightnesses (Cutri et
al. 2006). Because most of the galaxies presented in this work are
relatively bright, the typical uncertainties in the photometry are
$\sim$ 3\%. Because 
of the bright apparent magnitudes of these galaxies, the loss of light
coming from low surface brightness regions must be comparable to that
found by  Bell et al. (2003) for K$\sim$13.5 galaxies, which is $\sim$
0.1 magnitudes, and should not affect our conclusions.
 
The sub-mm data were presented by Vlahakis et al. (2005). The
850~$\mu$m fluxes were measured from the SCUBA maps choosing an aperture that
would optimize the signal to noise ratio, taking into account the
optical extent of the galaxies indicated by the Digitized Sky
Survey\footnote{The Digitized Sky Surveys were produced at the
Space Telescope Science Institute under U.S. Government grant
             NAGW-2166. The images of these surveys are based on photographic data
obtained using the Oschin Schmidt Telescope on Palomar Mountain and
the UK Schmidt Telescope. The plates were processed into the present
compressed digital form with the permission of these institutions.}.

The \iras ~fluxes were obtained by co-adding images 
using the NASA/IPAC Infrared Science Archive's IRAS Scan Processing
and Integration (SCANPI) tool and selecting the Noise-Weighted
Mean. Given the much higher quality
of the 24 $\mu$m fluxes from MIPS, we will not use the \iras ~25
$\mu$m fluxes in the analyses that follow.

\section{Analysis}

\begin{figure}
\vspace{60mm}
\includegraphics{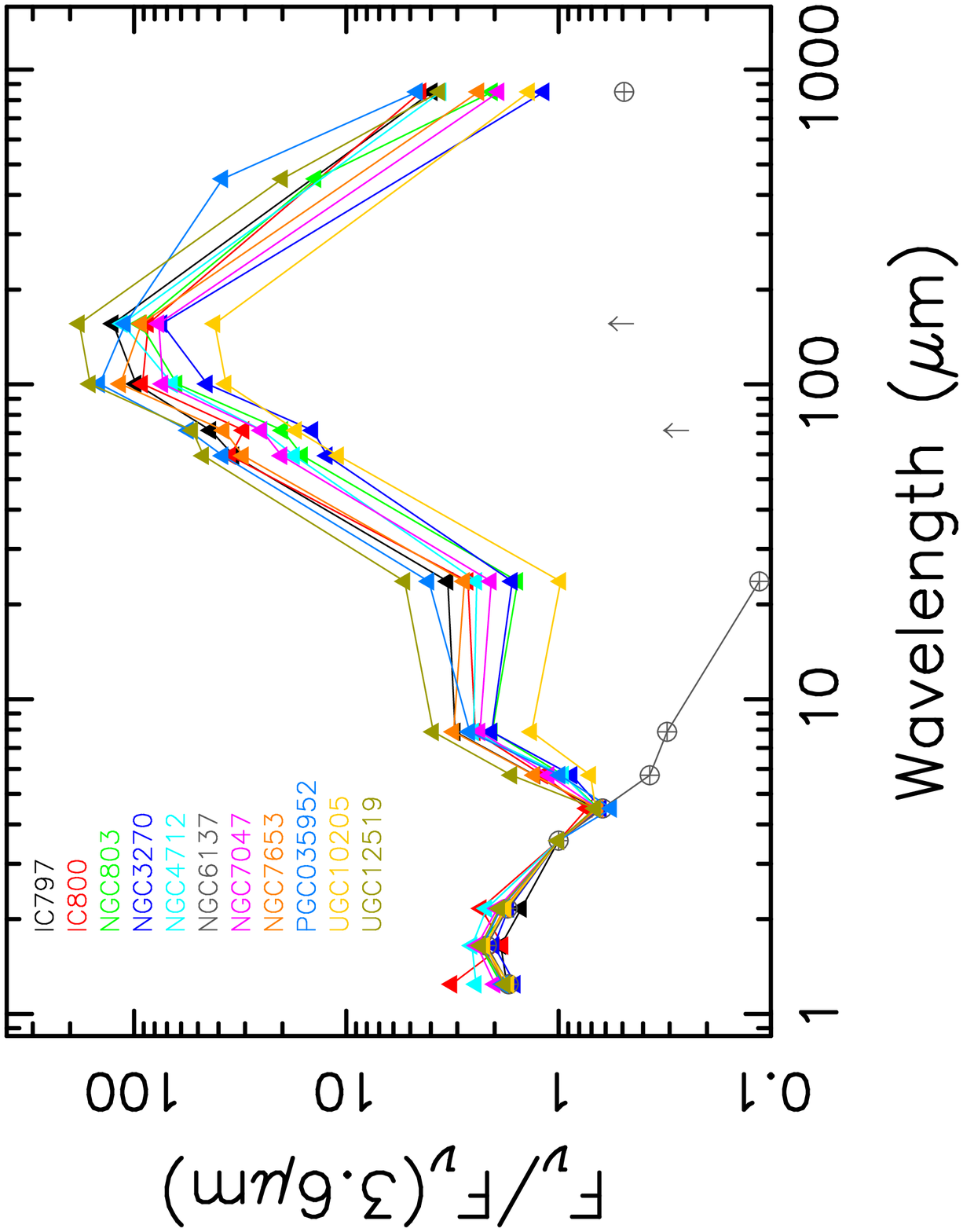}
\caption{Spectral energy distribution for the ensemble of SLUGS
  galaxies observed for this project, normalized at 3.6 $\mu$m. The
  stellar component shows little variation, while all galaxies, with
  the exception of NGC~6137 show a bump in the mid-IR  due to the aromatic
  features, followed by a strong peak in the far-IR caused by the
  presence of cold dust. For NGC~6137 the
  $Spitzer$ measurements at 70 and 160 $\mu$m are only upper
  limits. The combined 24 and 850 $\mu$m data points are
  suggestive of a power-law emission in the  mid- and far-IR for this galaxy.}
\end{figure}

Figure 1 presents a mosaic of all images taken with IRAC and MIPS
for the SLUGS galaxies observed with $Spitzer$. The images
are ordered in observed wavelength and the scale bar in each
panel represents a projected angular distance of 2 kpc. 
All galaxies are resolved in the IRAC bands, and different
subcomponents of galaxies are easily identified.
Similar to the SINGS galaxies (Dale et al. 2005) and the galaxies of
Hinz et al. (2006, 2007), the older stellar 
populations provide the main contribution in bluer IRAC bands (3.6 and
4.5 $\mu$m) while the dustier disk becomes more prominent in the redder
5.8 and 8.0 $\mu$m bands. Although most galaxies of this sample
are resolved in the MIPS bands, galaxy subcomponents are
only easily identified in the MIPS 24 $\mu$m images. While the galaxy
outlines and most prominent components can be seen in the 70$\mu$m
images, the 160 $\mu$m observations in general show very little
detail, though the extent to which the emission is seen is comparable
to that of the stellar component. 
One galaxy (NGC~6137) is only weakly detected at 70 and 160 $\mu$m and shows a
point-like source at 24 $\mu$m, probably due to the radio source at the
center of this galaxy (Giovanini et al. 2005). The new data
suggest that probably all, rather than $\sim$ 20\%, of the 850 $\mu$m
emission detected by Vlahakis et al. (2005) for this galaxy is due to an
active nucleus.

The spectral energy distribution of the SLUGS galaxies normalized at 3.6
$\mu$m, where the galaxy light is dominated by old stellar
populations, is shown in Figure 2. Each galaxy is represented by a
different color as noted in the key. Most objects show similar
spectral shapes, with intensities peaking at $\sim$ 100 $\mu$m. The
only exception is NGC~6137. 

\subsection{Dust component temperatures}

\begin{table*}
\scriptsize
\caption{Temperature fit parameters for SLUGS and SINGS galaxies}
\begin{center}
\begin{tabular}{lccrrrr}
\hline
\hline
Galaxy  & T$_{warm}$& T$_{cold}$ &M$_{cold}$/M$_{warm}$\cr
{}      &   K       &    K       &  {} \cr
\hline
IC~797       & 53.9$\pm$ 1.3 & 18.1$\pm$ 1.0 & 1213\cr
IC~800       & 51.1$\pm$ 2.8 & 16.2$\pm$ 2.8 & 1065\cr
NGC~3270     & 56.9$\pm$ 1.9 & 18.5$\pm$ 0.9 & 1918\cr
NGC~4712     & 56.2$\pm$ 0.9 & 17.0$\pm$ 0.4 & 3213\cr
NGC~7047     & 54.5$\pm$ 2.8 & 18.8$\pm$ 1.4 & 1248\cr
NGC~7653     & 54.7$\pm$ 4.0 & 19.9$\pm$ 2.1 & 1008\cr
NGC~803      & 54.7$\pm$ 1.0 & 18.4$\pm$ 0.4 & 1812\cr
PGC~035952   & 54.7$\pm$ 4.4 & 19.0$\pm$ 2.5 & 1054\cr
UGC~10205    & 52.5$\pm$ 2.8 & 18.4$\pm$ 1.8 & 1098\cr
UGC~12519    & 56.0$\pm$ 2.1 & 19.6$\pm$ 1.0 & 1054\cr
    NGC~337  & 52.8$\pm$ 3.2 & 20.6$\pm$ 3.8 &   337\cr
    NGC~1482  & 56.1$\pm$ 3.6 & 23.5$\pm$ 5.2 &   109\cr
    NGC~2798  & 58.0$\pm$ 4.3 & 24.4$\pm$ 5.4 &   116\cr
    NGC~2976  & 54.9$\pm$ 3.7 & 19.1$\pm$ 5.1 &   800\cr
    NGC~3190  & 54.2$\pm$ 3.4 & 19.9$\pm$ 3.9 &   766\cr
      Mrk~33  & 60.7$\pm$ 5.1 & 24.3$\pm$ 7.2 &   111\cr
    NGC~3521  & 55.2$\pm$ 2.9 & 18.5$\pm$ 3.8 &   985\cr
    NGC~3627  & 55.2$\pm$ 3.0 & 19.2$\pm$ 4.6 &   571\cr
    NGC~4536  & 55.5$\pm$ 2.9 & 20.6$\pm$ 4.6 &   225\cr
    NGC~4569  & 58.4$\pm$ 3.4 & 19.7$\pm$ 2.9 &   973\cr
    NGC~4631  & 52.6$\pm$ 3.4 & 18.9$\pm$ 4.0 &   539\cr
    NGC~4826  & 52.4$\pm$ 3.8 & 20.3$\pm$ 5.0 &   362\cr
    NGC~5195  & 59.4$\pm$27.9 & 25.7$\pm$ 9.5 &   258\cr
    NGC~5713  & 54.9$\pm$ 2.9 & 20.2$\pm$ 4.1 &   307\cr
    NGC~5866  & 57.2$\pm$16.1 & 24.5$\pm$ 3.1 &   878\cr
    NGC~7331  & 54.4$\pm$ 3.2 & 19.4$\pm$ 2.8 &  660\cr
\hline
\end{tabular}
\end{center}
\end{table*}

The new MIPS data combined with existing \iras ~and SCUBA measurements
can be used to estimate the dust component temperatures and masses
using composite grey-body curves (e.g., Vlahakis et
al. 2005). For two components these are:
\begin{equation}
 F_\nu =  A_w \nu^\beta B_\nu(\frac{\nu}{1+z},T_w) +  A_c
 \nu^\beta B_\nu(\frac{\nu}{1+z},T_c)
\end{equation} 
where A$_w$ and A$_c$ are the relative contributions due to the warm and
cold dust components. The masses can be estimated from (Vlahakis et al. 2005):
\begin{equation}
M = \frac { F_\nu D^2 } {k_d} \left\lbrack  \frac { A_w }
{B_\nu(\frac{\nu}{1+z},T_w)} + \frac {A_c}{B_\nu(\frac{\nu}{1+z},T_c)}\right\rbrack \frac {1}
{ A_w + A_c},
\end{equation}
where $k_d$ = 0.07 $\pm$0.02 m$^2$ kg$^{-1}$ is the average dust mass
opacity measured at 850 $\mu$m  by James et al. (2002), $D$ is the
galaxy distance and $B_\nu$ the modified Planck function.
Since the average radiation field combines emission at several
intensity levels (e.g., Draine \& Li 2007; Draine et al. 2007; see below), the
temperatures should only be taken as a rough estimate of the
``effective dust temperature'' (Blain et al. 2004).
The fits were calculated considering three values for the dust emissivity index
($\beta$ = 1.0, 1.5, 2.0), $\beta$ = 2.0 producing better constrained
fits. This is also the value of $\beta$ used  by Vlahakis et al. (2005)
in their analysis. The fits are presented in Figure 3, while Table 5
lists the temperatures and errors for the cold and warm components,
and the estimated mass ratio between warm and cold components.
The larger wavelength range of the present sample allows more
constrained fits than was possible for Vlahakis et al. (2005). For
the 3 objects in common for which they calculate two-component fits,
the most dramatic difference is found for 
UGC~12519, where the measurements change from ($T_w$, $T_c$,
$M_{cold}/M_{warm}$) = ( 28, 18, 12) to (56.0, 19.6, 1054). For
NGC~803 Vlahakis et al. (2005) find two 
possible solutions (33, 18, 92) and (60, 19, 2597). The latter is
closer to the values measured here (54.7, 28.4,
1812). For PGC~35952 the $IRAS$+SCUBA measurements produce (58, 18,
1859), while we find (54.7, 19.0,1054). The
temperatures agree within the estimated errors, though the dust mass
ratios can differ significantly. As noted by Draine et al. (2007) even
small shifts in temperatures imply very large changes in the dust
masses, as seen in this comparison.
Because of the paucity of data longwards of 160 $\mu$m for the SLUGS
sample analysed here (only 3 galaxies have 450 $\mu$m data) we cannot
test for the existence or absence of dust colder than 18 K which has
been found in low-metallicity dwarf galaxies (e.g. Galliano et al. 2005). 

\begin{figure*}
\vspace{190mm}
\includegraphics{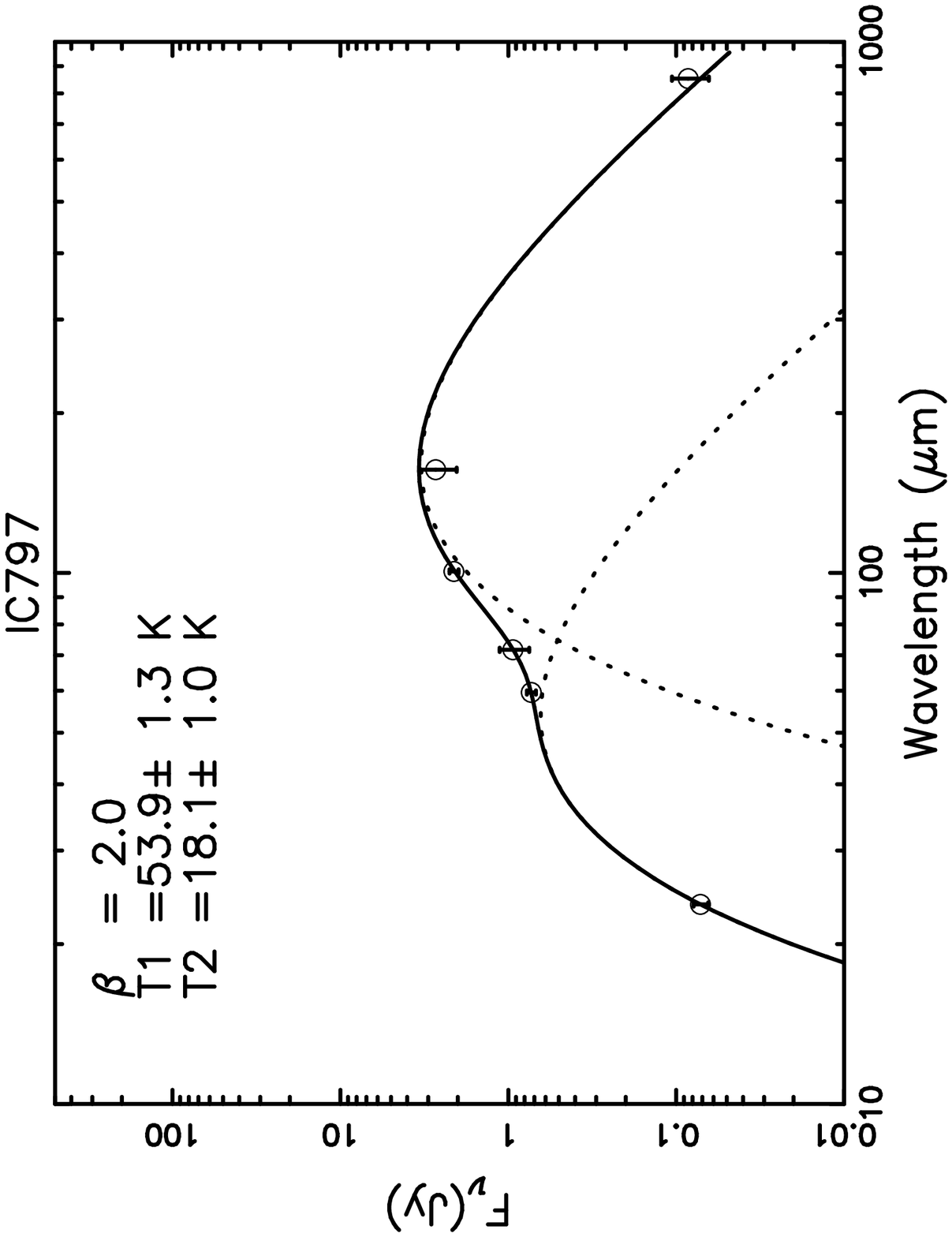}
\includegraphics{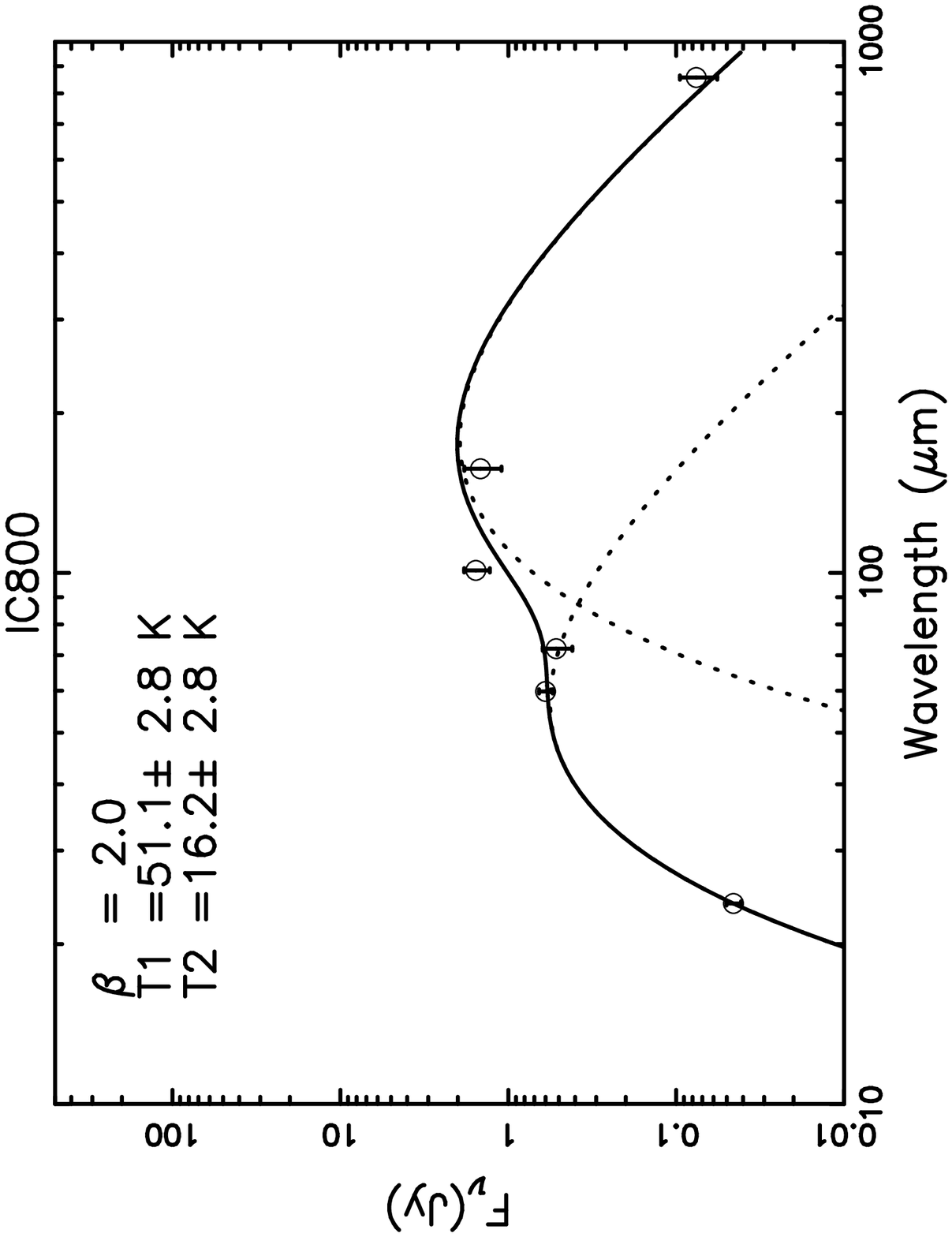}
\includegraphics{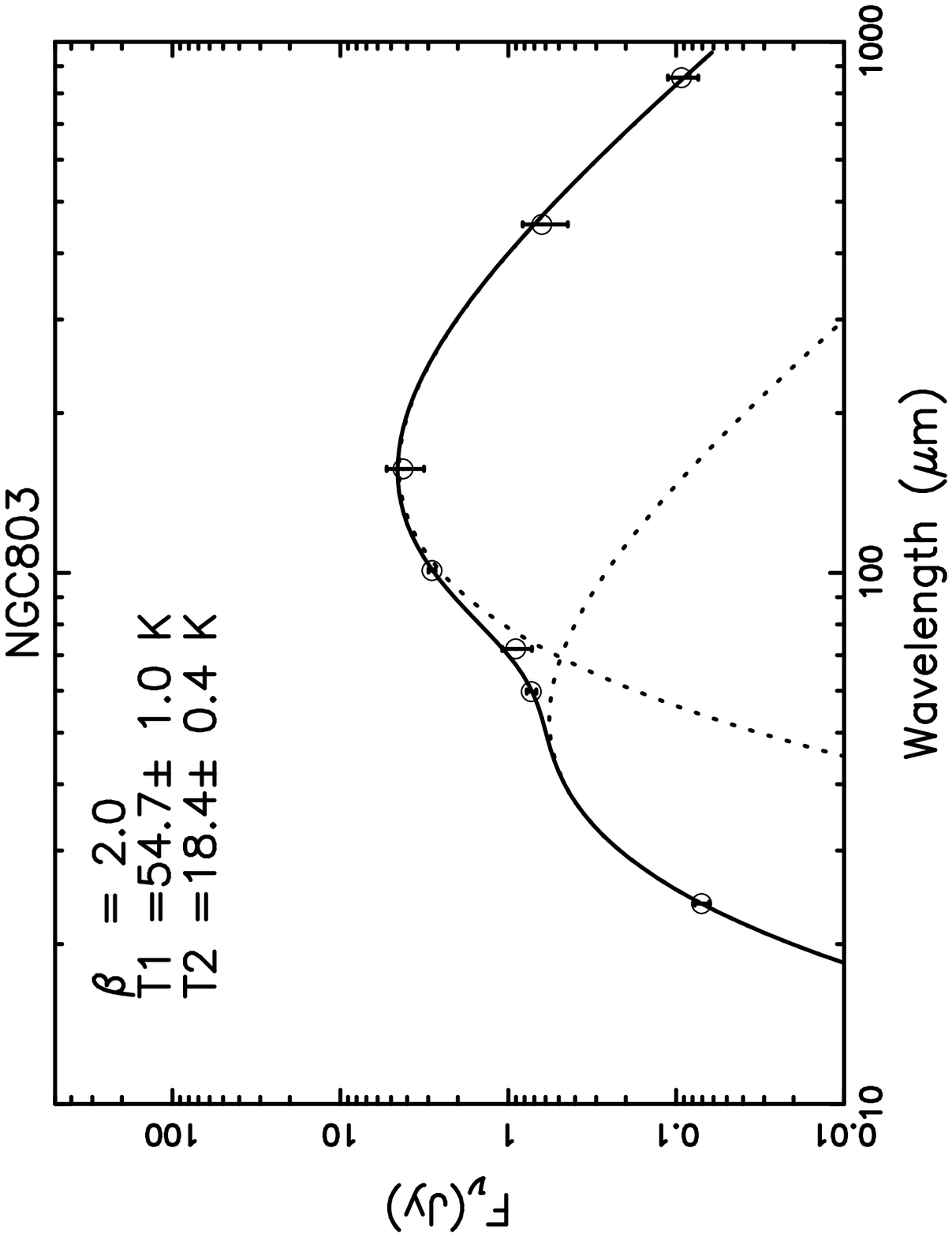}
\includegraphics{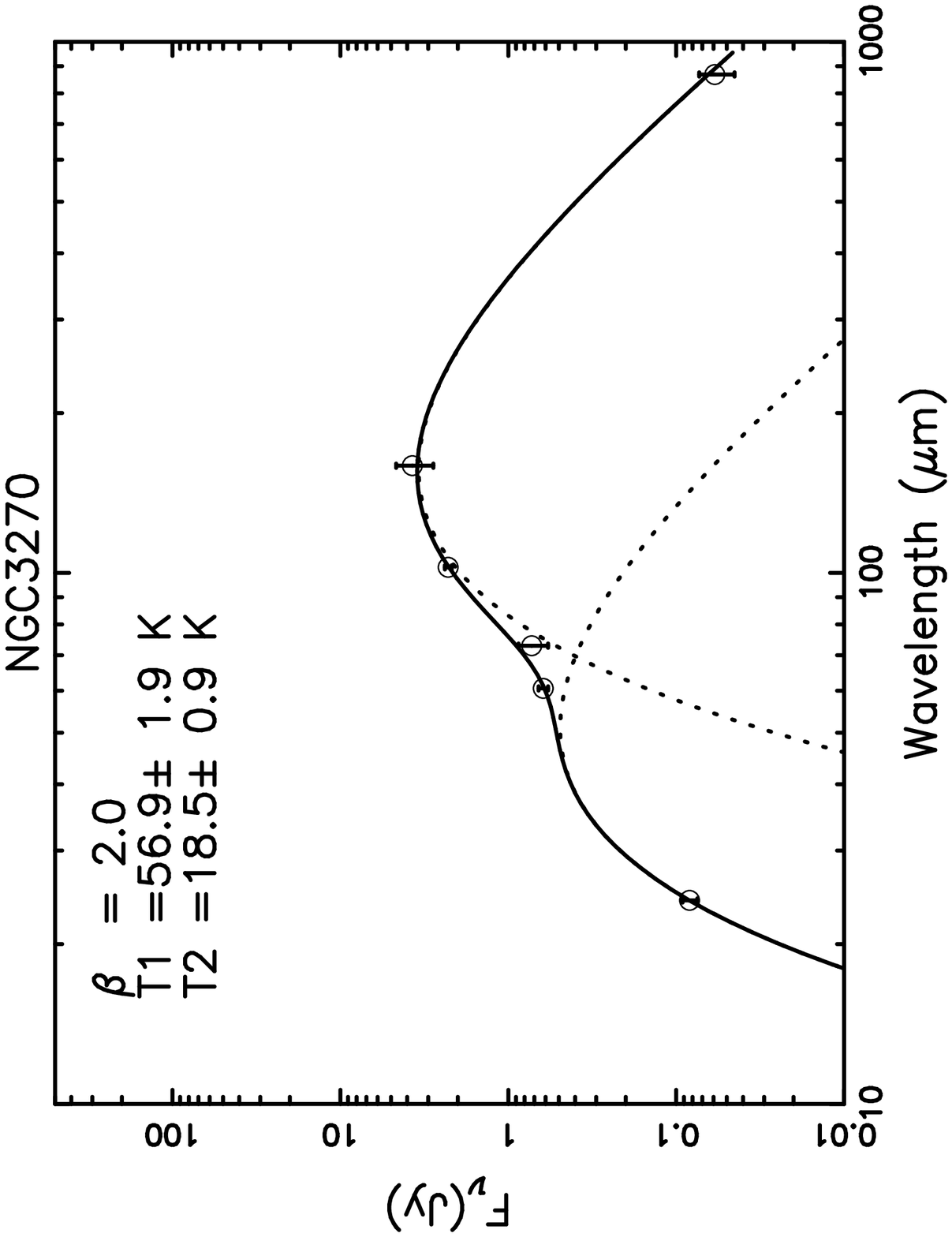}
\includegraphics{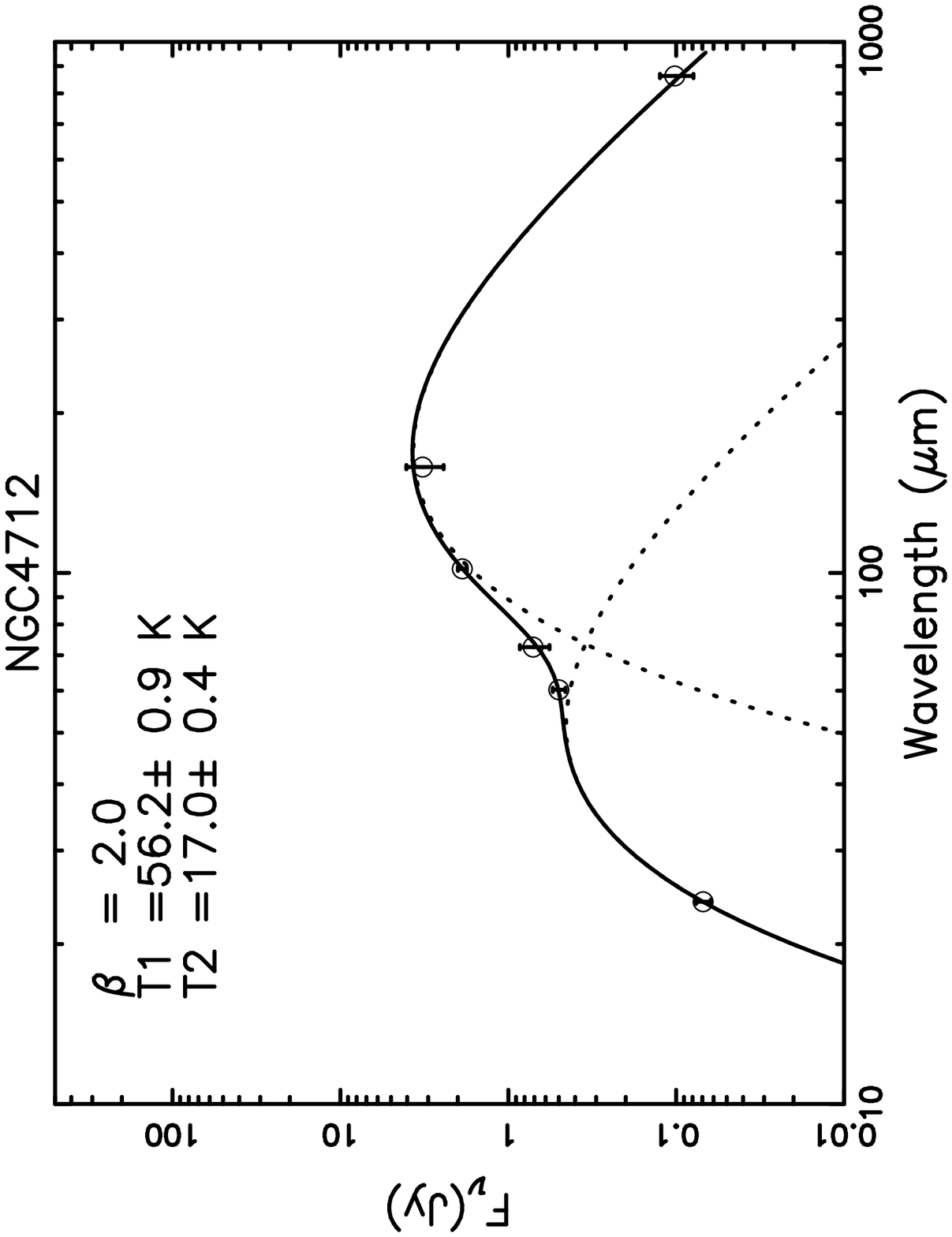}
\includegraphics{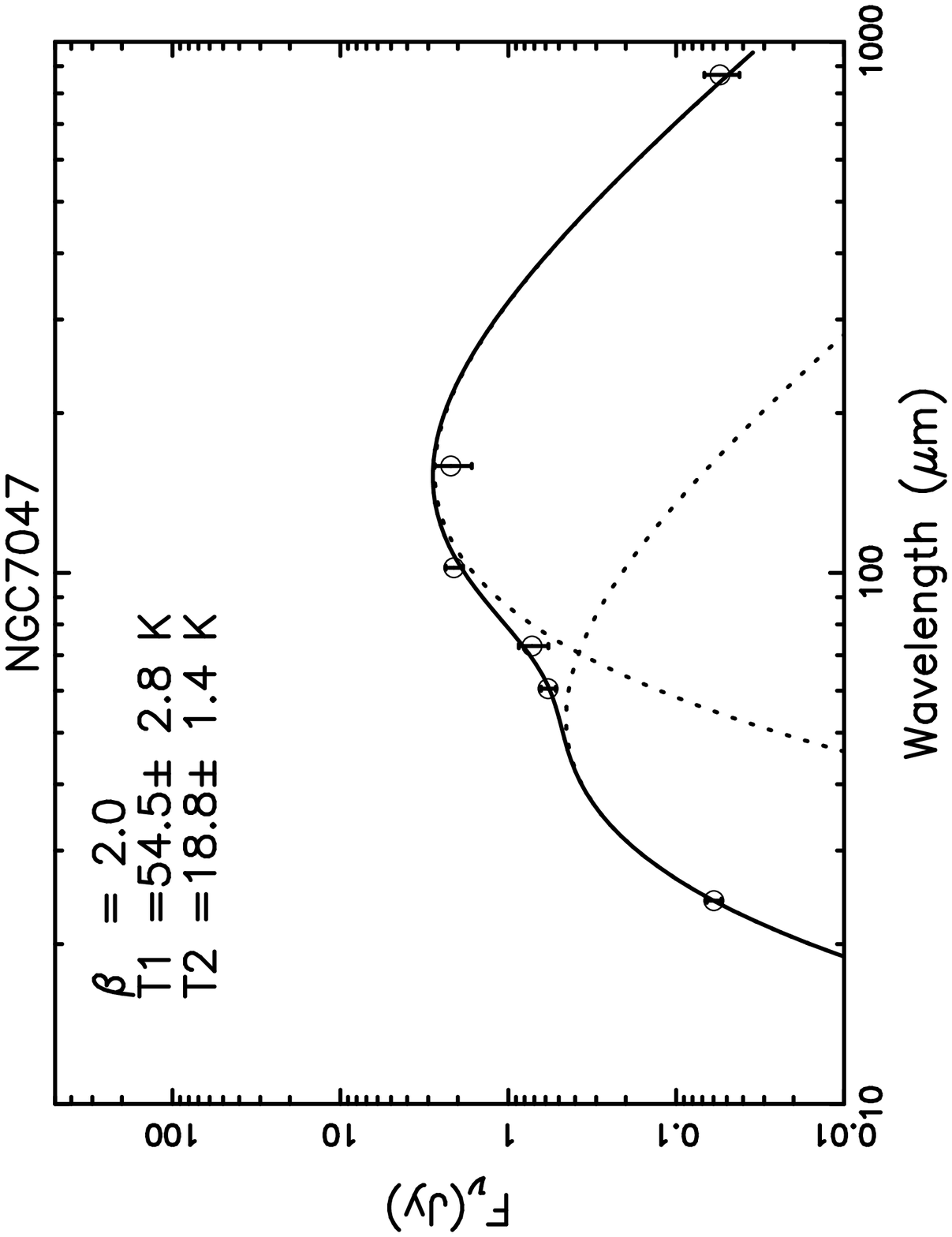}
\includegraphics{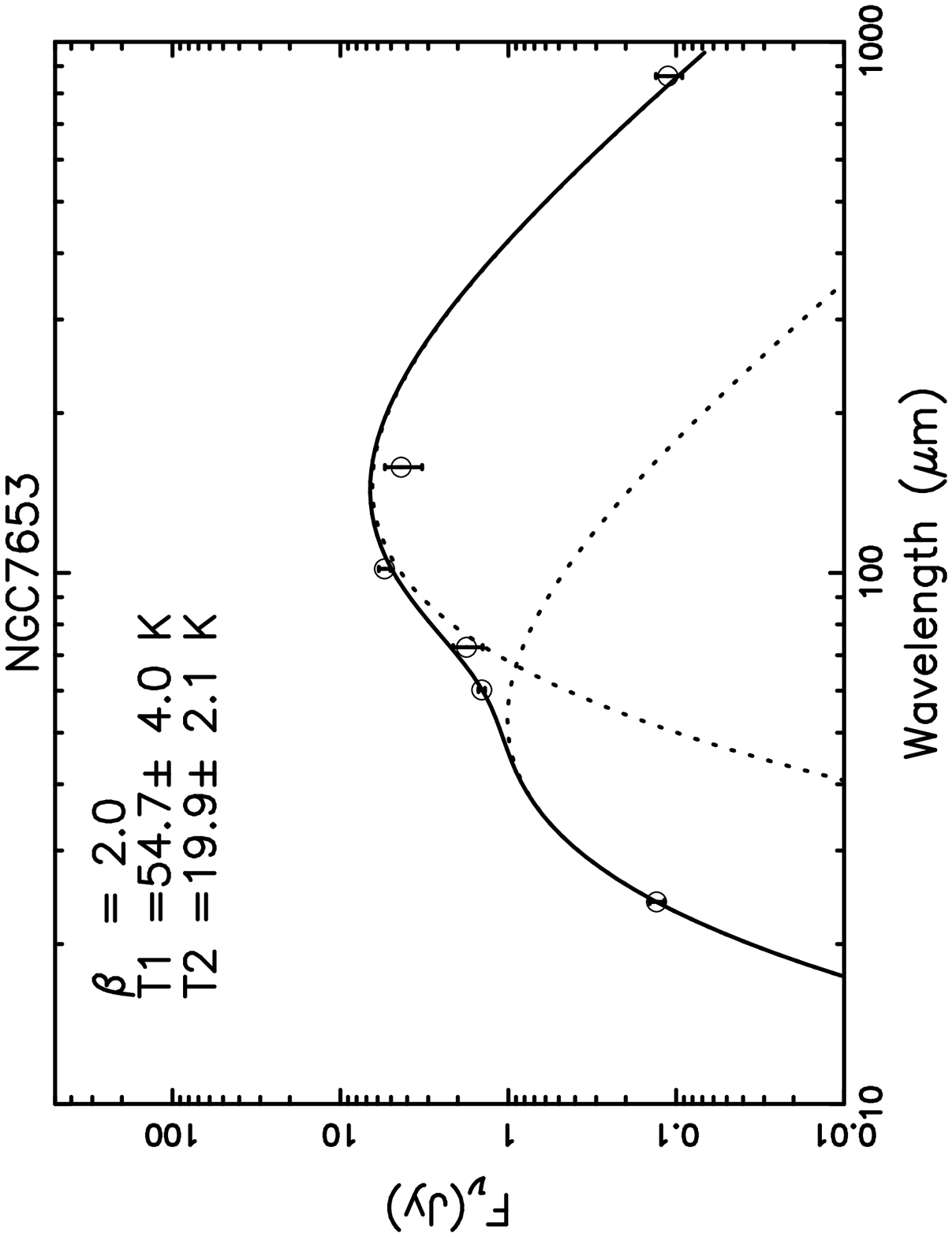}
\includegraphics{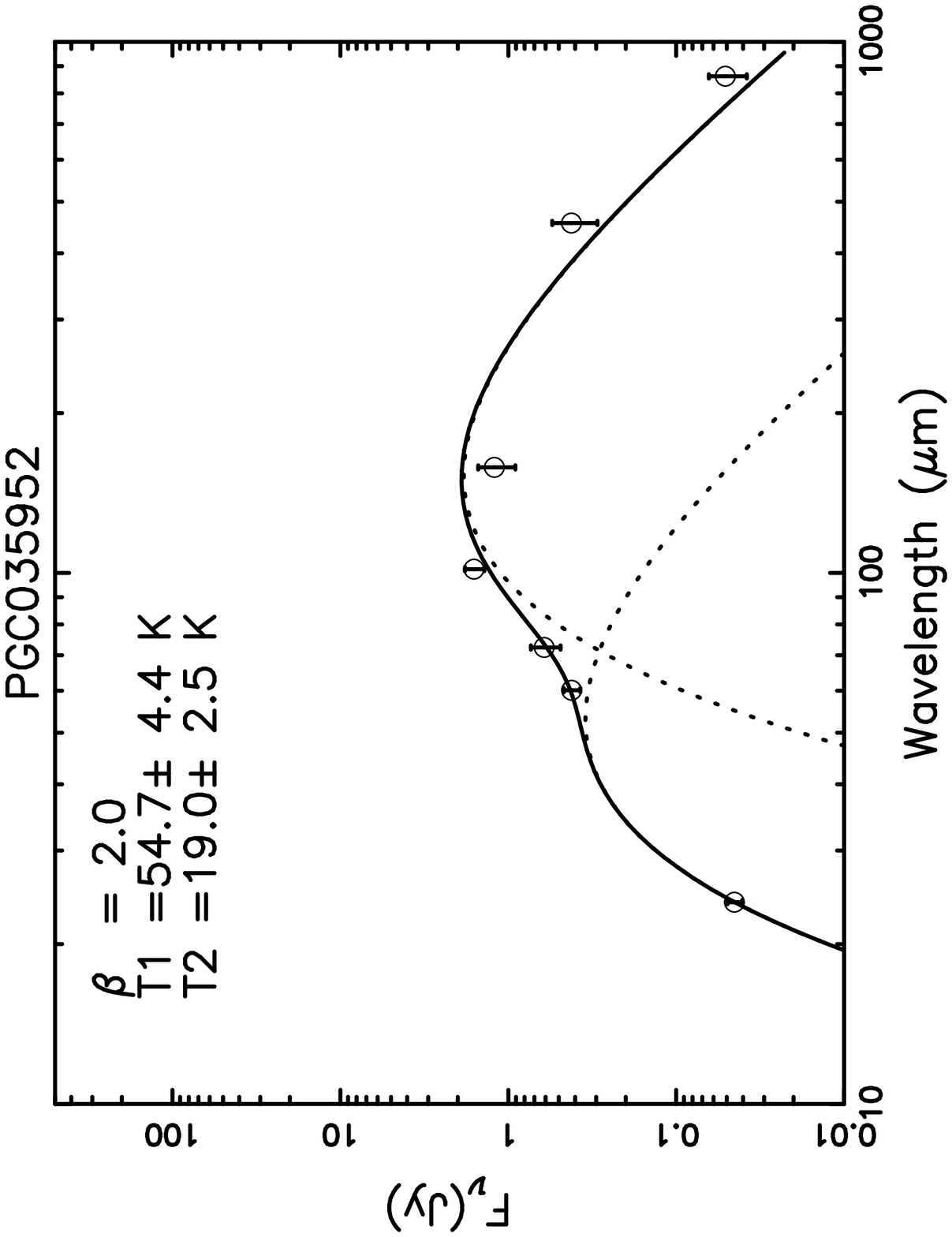}
\includegraphics{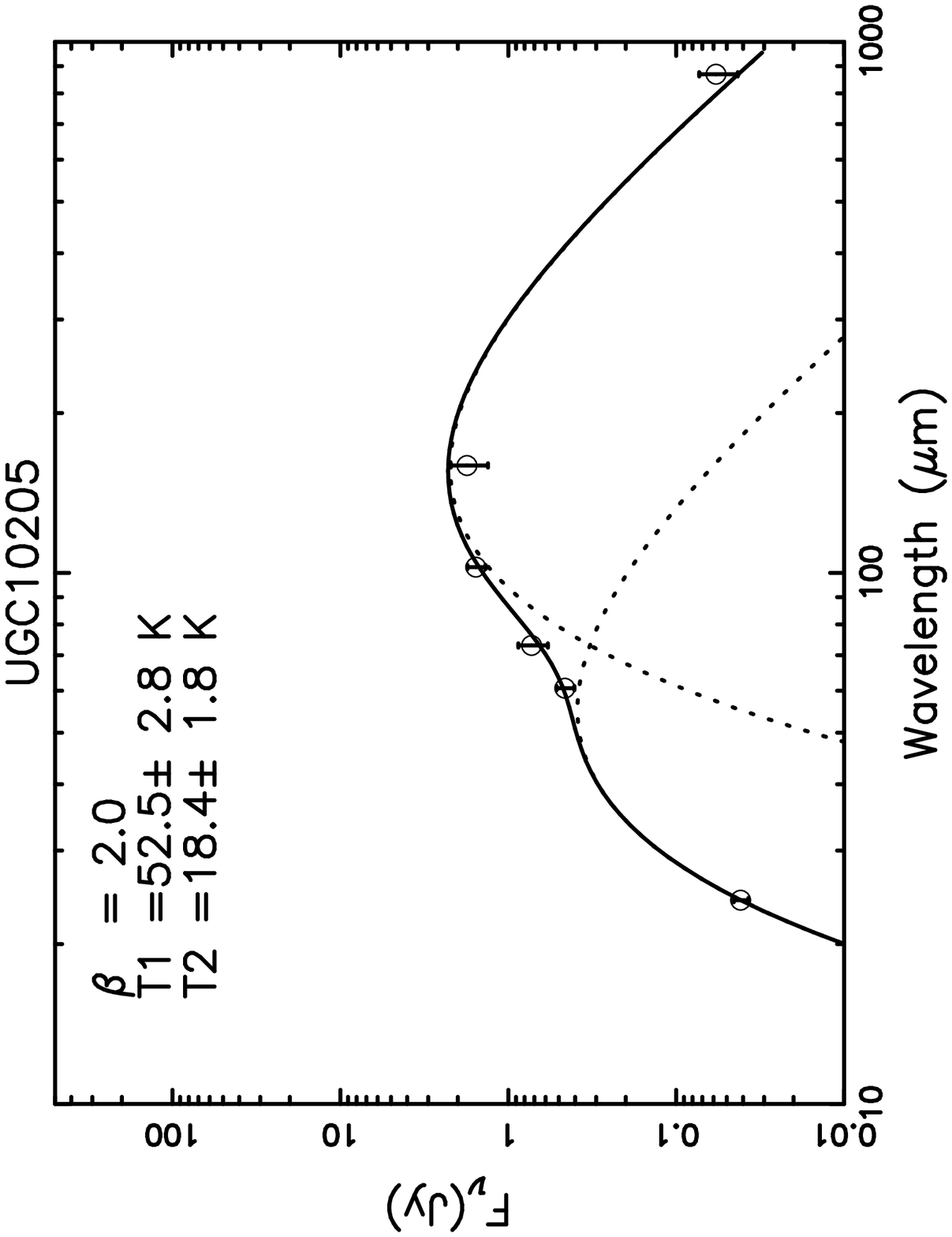}
\includegraphics{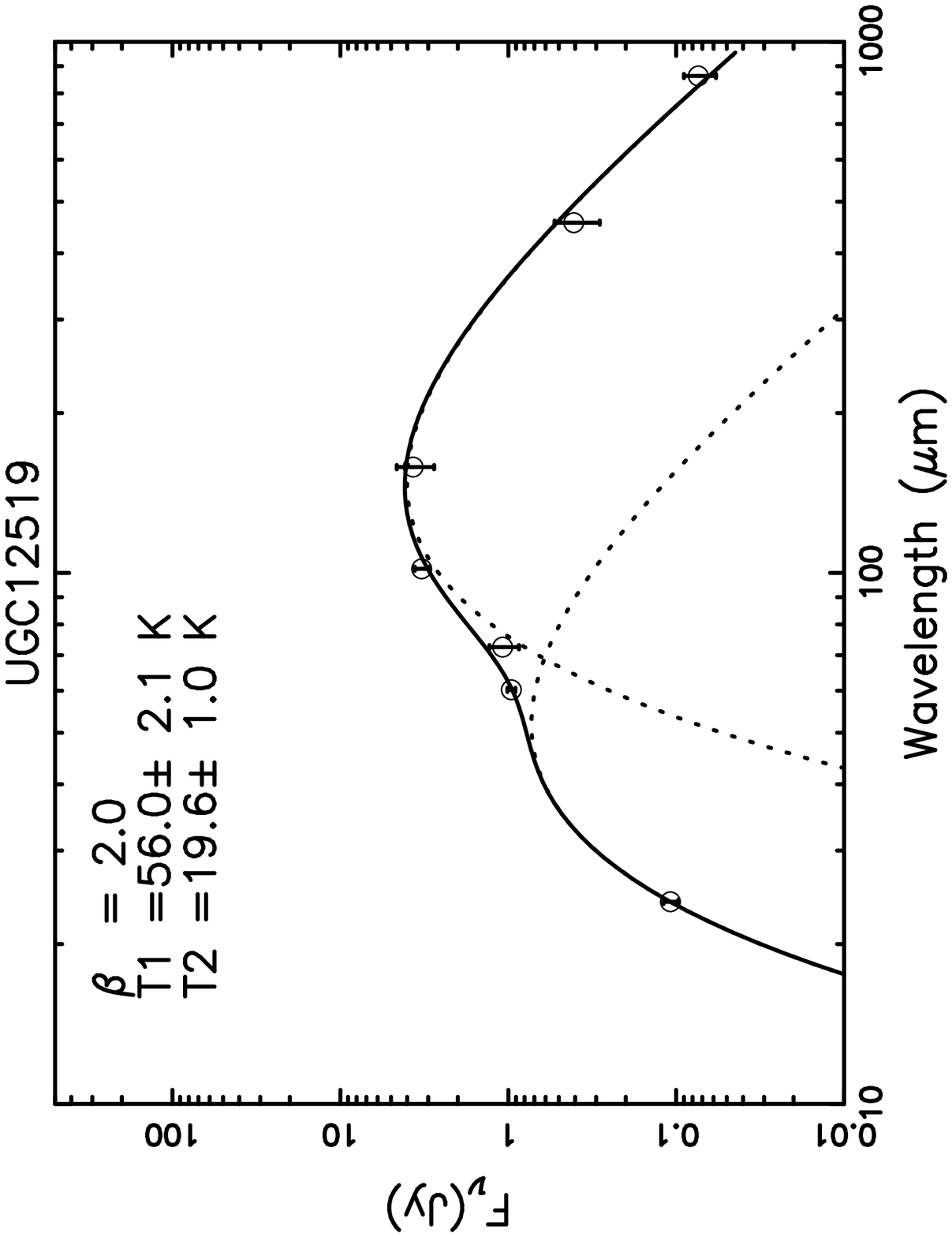}
\vskip 1cm
\caption{Temperature fits using modified blackbody curves with the
  emissivity index $\beta$=2, assuming two temperature components. The
  dotted curves represent the individual components, the solid curve
  the combined flux. The values of best fit temperatures and
  associated errors are also shown in the figure.
}
\end{figure*}

\subsection{Galaxy models}

\begin{table*}
\scriptsize
\caption{Results using Draine \& Li dust models fitting $U_{max}$}
\begin{center}
\begin{tabular}{lrrrrrr}
\hline
\hline
Galaxy  & M$_{dust}$ & L$_{dust}$ & $\gamma$~~~ & $<$U$>$ & $q_{PAH}$& $U_{max}$ \cr
 {}     & $M_{\odot}$ & $L_{\odot} $ & \%    & {}      & \% & {} \cr
\hline
IC~797& 6.71& 8.85& 7.11& 1.01& 4.58    & 10$^{3}$\cr
IC~800& 7.51& 9.44&15.55& 0.63& 3.90    & 10$^{3}$\cr
NGC~3270& 8.33&10.57& 0.55& 1.28& 4.58  & 10$^{6}$\cr
NGC~4712& 8.14&10.17& 1.20& 0.79& 3.90  & 10$^{5}$\cr
NGC~7047& 8.00&10.37& 1.94& 1.73& 4.58  & 10$^{4}$\cr
NGC~7653& 8.05&10.45& 4.55& 1.88& 4.58  & 10$^{3}$\cr
NGC~803& 7.43& 9.64& 0.00& 1.20& 4.58   & 10$^{6}$\cr
PGC~035952& 7.77& 9.85& 8.57& 0.88& 2.50 & 10$^{4}$\cr
UGC~10205& 8.18&10.38& 7.33& 1.16& 3.90 & 10$^{3}$\cr
UGC~12519& 7.95&10.31& 1.10& 1.71& 3.90 & 10$^{6}$\cr
\hline

\hline
\end{tabular}
\end{center}
\end{table*}

\begin{table*}
\scriptsize
\caption{Results using Draine \& Li dust models with fixed $U_{max}$=10$^{6}$}
\begin{center}
\begin{tabular}{lrrrrr}
\hline
\hline
Galaxy  & M$_{dust}$ & L$_{dust}$ & $\gamma$~~~ & $<$U$>$ & $q_{PAH}$\cr
 {}     & $M_{\odot}$ & $L_{\odot} $ & \%    & {}      & \% \cr
\hline
IC~797& 6.57& 8.82& 0.74& 1.31& 4.58\cr
IC~800& 7.05& 9.40& 0.73& 1.64& 4.58\cr
NGC~3270& 8.33&10.57& 0.55& 1.28& 4.58\cr
NGC~4712& 8.14&10.17& 0.82& 0.78& 3.90\cr
NGC~7047& 8.02&10.36& 0.65& 1.62& 4.58\cr
NGC~7653& 7.98&10.44& 0.49& 2.12& 4.58\cr
NGC~803& 7.43& 9.64& 0.00& 1.20& 4.58\cr
PGC~035952& 7.51& 9.81& 1.86& 1.48& 2.50\cr
UGC~10205& 7.91&10.36& 0.18& 2.04& 4.58\cr
UGC~12519& 7.95&10.31& 1.10& 1.71& 3.90\cr
\hline
\end{tabular}
\end{center}
\end{table*}

More robust estimates of the (total) dust masses can be obtained using
models with a more complex dust composition
(e.g., D\'esert et al. 1990; Guiderdoni
et al. 1998; Dale et al. 2001; Dale \& Helou 2002; Marshall et
al. 2006; Siebenmorgen \& Kr\"ugel 2007; Draine \& Li, 2007; Draine et
al. 2007).
These models, combined with
the empirical SEDs constructed using existing observations (e.g.,
Chary \& Elbaz 2001; Lagache et al. 2003; Lagache et al. 2004; Rieke
et al. 2009)  have been used to explain the strong evolution
as a function of redshift in number and density that is observed in
the mid and far infrared (e.g., Puget et al. 1995; Le Floc'h et
al. 2005; P\'erez-Gonz\'alez et al. 2008).  Typically the models
describe the mid- and far-IR properties of galaxies combining 
the UV output from hot young stars, which is re-radiated by small size
particles, and larger grains that are heated stochastically or are in
thermal equilibrium (D\'esert, Boulanger \& Puget 1990).

To analyse the sample presented in this paper and compare their
properties with those measured for SINGS galaxies, we follow the
analysis of  Draine et al. (2007), by using the models calculated by
Draine \& Li (2007). The latter work builds upon many previous papers by
Draine and collaborators (e.g., Li \& Draine 2001; Weingartner \&
Draine 2001a; 2001b) by taking into account the
results of recent observations (e.g., Smith et al. 2007). 
These models are based on a series of physically motivated assumptions,
and can be used to estimate the galaxies' total dust masses.
The dust models are parameterized both by the dust composition and by the stellar
radiation field incident on the dust grains. 
In the Draine et al. (2007) models, most of the dust is assumed to
be located in a diffuse ISM that is heated by a general
interstellar radiation field (e.g., Lonsdale \& Helou 1987) typical of
normal spiral galaxies ($U_{min}$). This field has an equilibrium temperature
$\sim$ 16.2 K (based on Milky Way measurements by Finkebeiner, Davis
\& Schlegel 1999) while a range of radiation field intensities
heats the dust particles to different temperatures.
The fraction of galaxy dust that is 
subject to the intense radiation field generated by
hot young stars ($U_{max}$) is denoted by $\gamma$ and is quantified as a
percentage of the total dust emission of the galaxy.
The emission at $\lambda$ $<$ 20 $\mu$m (Draine \& Li 2007) is
mainly due to aromatic grains, whose contribution is characterized by the
$q_{PAH}$ parameter, which quantifies the ratio between the mass in
aromatic grains containing less than 1000 carbon atoms and the total dust mass.
Weingartner \& Draine (2001) calculated several models that use different
grain size distributions and with different $q_{PAH}$ values, but
are constrained to follow the average ISM extinction law, reproducing
the Milky Way's wavelength-dependent extinction curves (Draine \& Li
2007). The models, described in Table 3 of Draine et al. (2007),
show slight variations in the total dust mass to hydrogen ratio
but have values of $M_{dust}/M_H \sim 0.01$, slightly higher
than the prefered value of Draine et al. (2007) of 0.0073, calculated
assuming an approximately solar metallicity for the Galaxy's ISM.
The stellar continuum for wavelengths greater than 5 $\mu$m is well
described by a 5000 K black-body (e.g., Smith et al. 2007; Draine et
al. 2007). Following Draine et al. (2007) the SEDs of SLUGS galaxies
are fit by minimizing  
\begin{equation}
\chi^2 = \sum_{i=1}^{N_{bands}} \frac { (f_{obs}(\nu_i) -
  F_{model}(\nu_i))^2} { err_{obs}^2 + err_{model}^2}.
\end{equation}
where, using the Draine et al. (2007) notation, the observed flux is a
combination of the stellar component with the emission due to the diffuse
ISM and dust heated by hot stars:
\begin{eqnarray}
F_{model}&&(\nu_i) = \Omega_* B_\nu(T_*, \nu_i) + \frac {M_{dust}}
{4\pi D^2} \times \nonumber \\
& & \left\lbrack(1-\gamma) p_\nu^0(j_m, U_{min}) + 
   \gamma p_\nu ( j_m, U_{min}, U_{max}, \alpha)\right\rbrack;\nonumber \\
\end{eqnarray}
$p_\nu^0$ and  $p_\nu$ represent the power radiated by the dust mixture per unit
mass and frequency that is subject to radiation field intensity of
$U_{min}$ ($p_\nu^0$) or to a range of intensities ($p_\nu$), assumed
to follow a power-law characterized by the exponent $\alpha$ (Draine
et al. 2007). Following these authors, we adopt $\alpha$=2.
The models for the PAH abundance, the radiation field, the
total $M_{dust}/M_H$ ratio and the extinction are combined into $j_m$,
the specific power radiated per unit dust mass\footnote[6]{Tables with
$j_m$ values are available at the
  {\url{ftp://ftp.astro.princeton.edu/draine/dust/irem4/}} site.}. 

When calculating the fits, we first normalize the
stellar component ($\Omega_*$) using the 2MASS photometry in $J$ and
$H$. This is followed by the calculation of the dust model normalization;
final fits are obtained by varying $\gamma$ and looping through all
models sharing the same proportion of aromatic masses ($q_{PAH}$) and
general radiation field ($U_{min}$). The results of this procedure are
presented in Table 6, while Figure 4 shows the measured SEDs and the
best fitting model. Also shown in Fig. 4 are the
graphic estimates of the stellar flux, the flux due to dust in
photo-dissociation regions and the flux emitted by
dust in ``the diffuse ISM heated by a single radiation field'' (Draine
et al. 2007). 
Most galaxies are reasonably well fit by the cold temperature diffuse
component with a small contribution from hot dust in
photo-dissociation regions (PDR). In the case of NGC~803 the fits are
consistent with no presence of PDRs, though the IRAC data do not show
any gross inconsistency with the presence of aromatic emission, which is
generally produced in regions with strong UV emission.  
The analysis by Bendo et al. (2008) of the correlation observed between the
aromatic emission at 8 $\mu$m and 160 $\mu$m in a sample of nearby spiral
galaxies as well as that of Haas, Klaas \& Bianchi (2002) shows that such
emission is expected. No fits are presented 
for NGC6137 because its FIR spectrum is probably dominated by
non-thermal emission.
The results in Table 6 also fit for $U_{max}$, which
represents the upper cut off of the stellar intensity field (Draine et
al. 2007). Fixing the value of  $U_{max}$ = 10$^{6}$ produces the fits
presented in Table 7. When fitting for $U_{max}$ we find that the residuals
in the far-IR are slightly smaller.

\begin{figure*}
\vspace{190mm}
\includegraphics{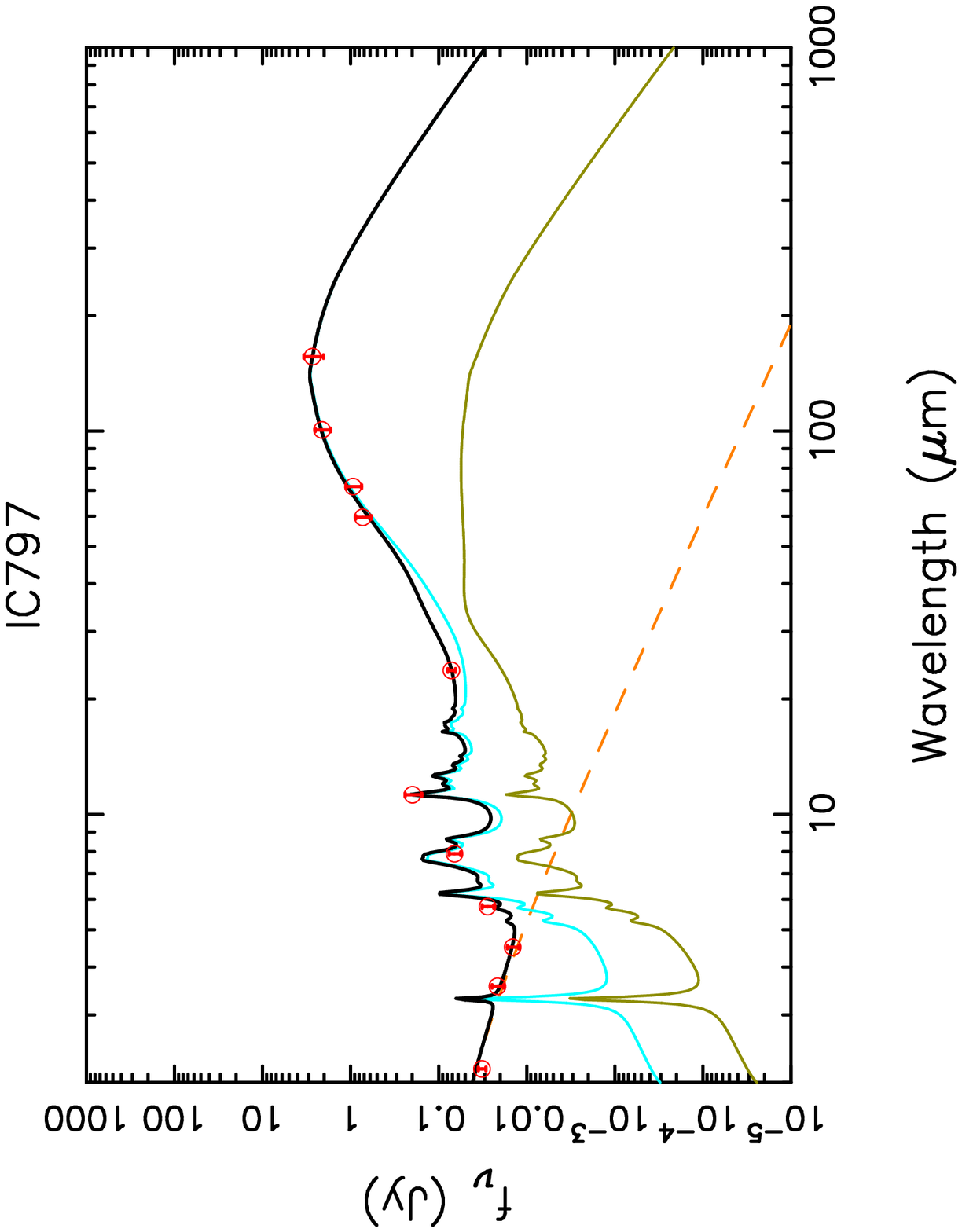}
\includegraphics{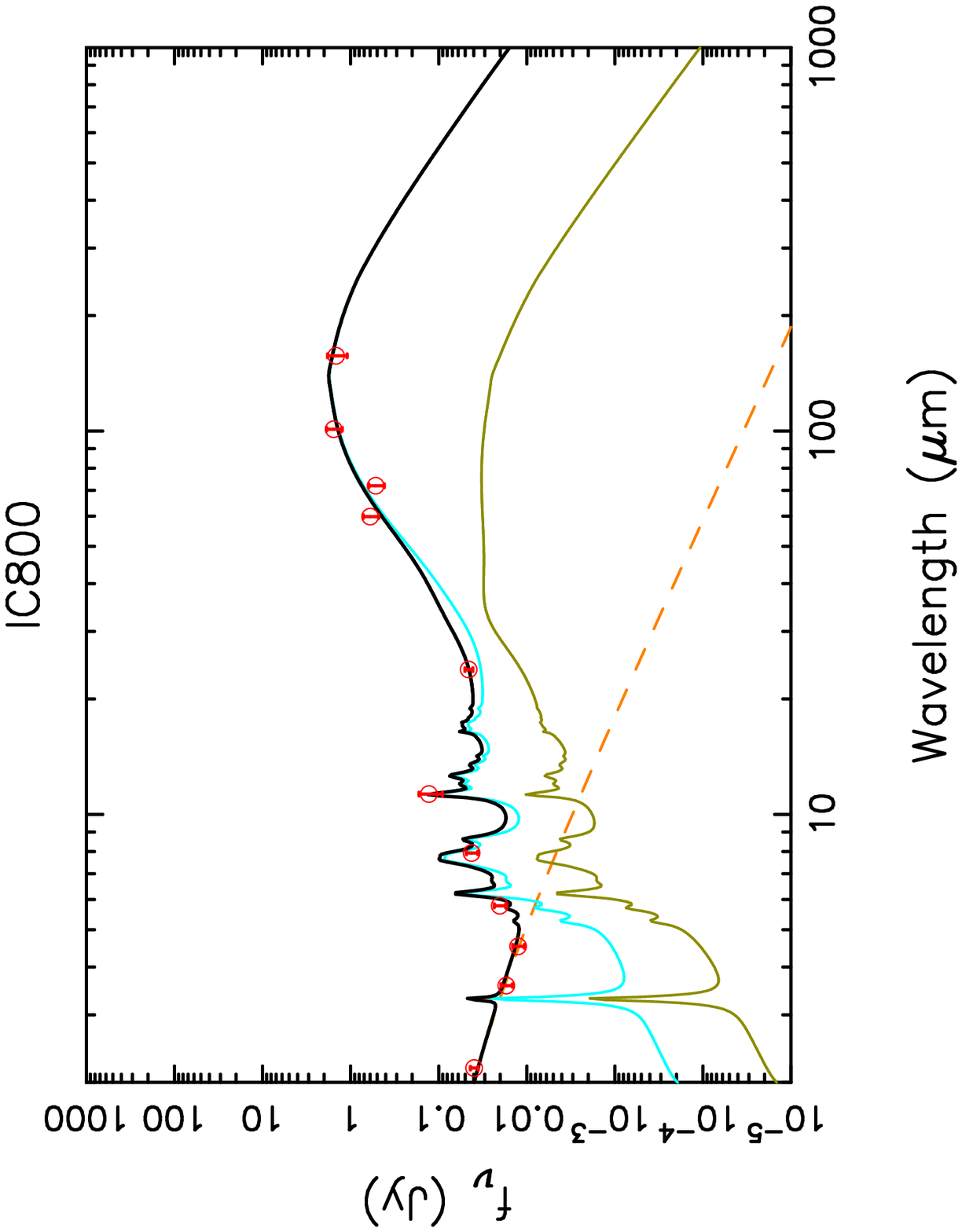}
\includegraphics{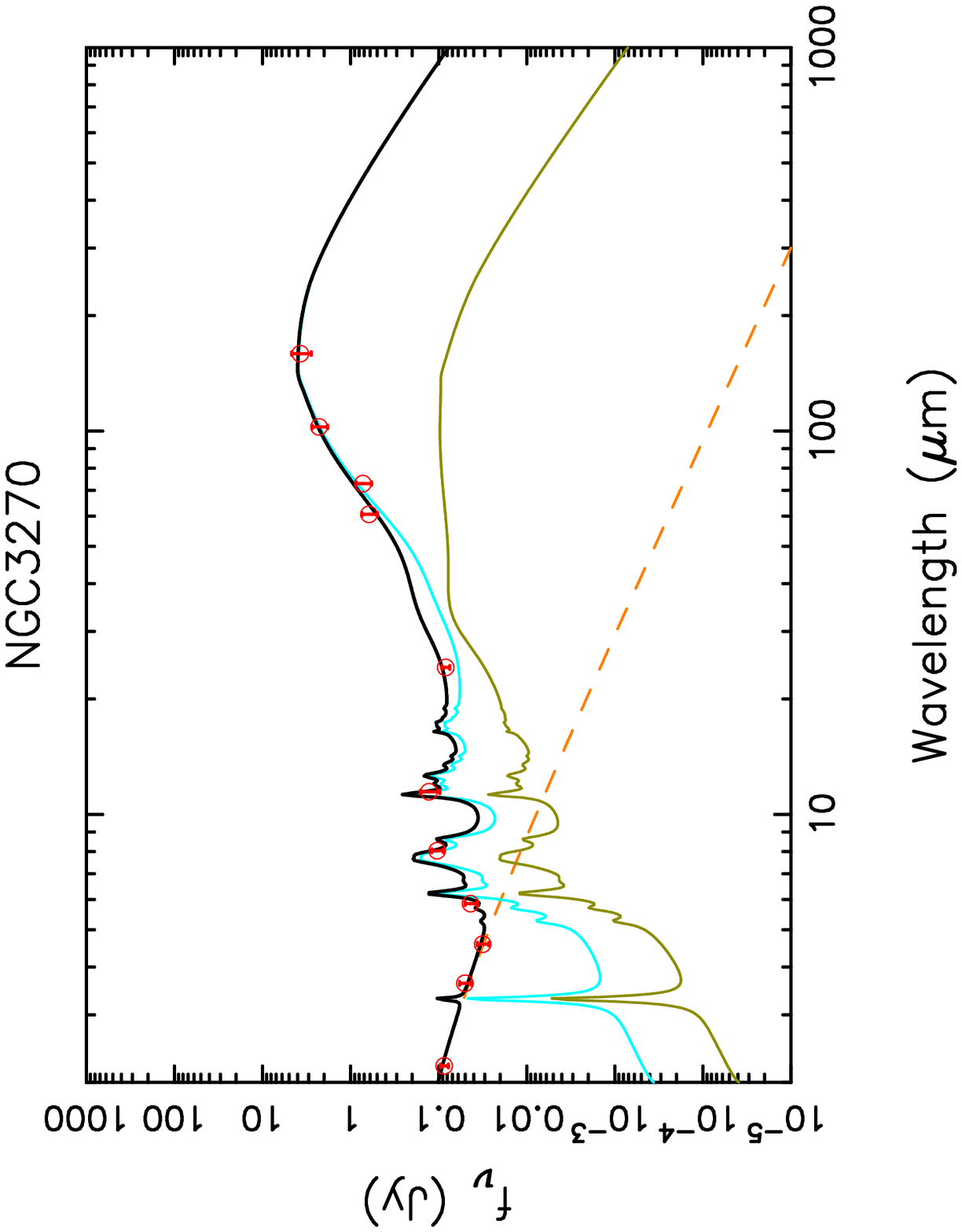}
\includegraphics{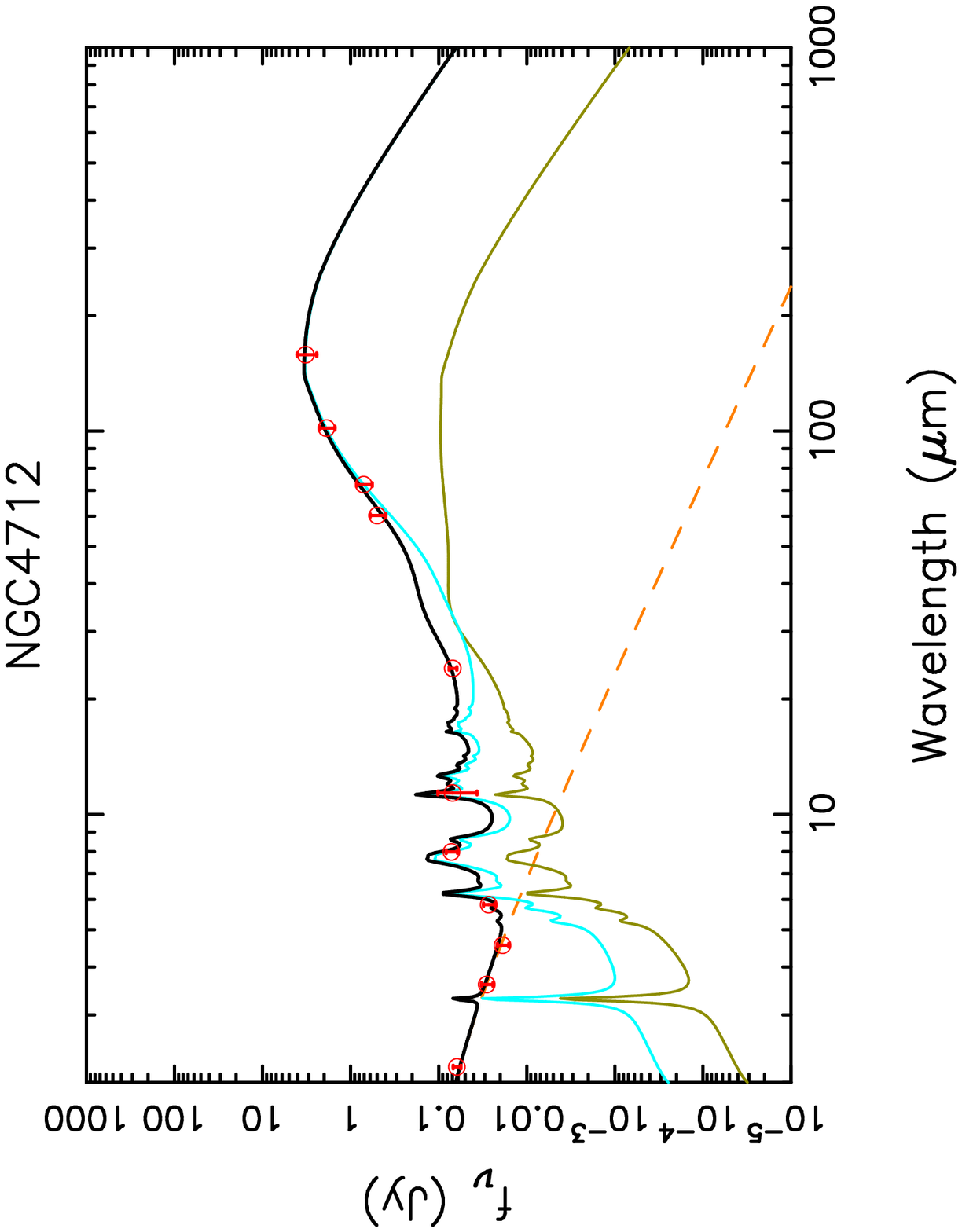}
\includegraphics{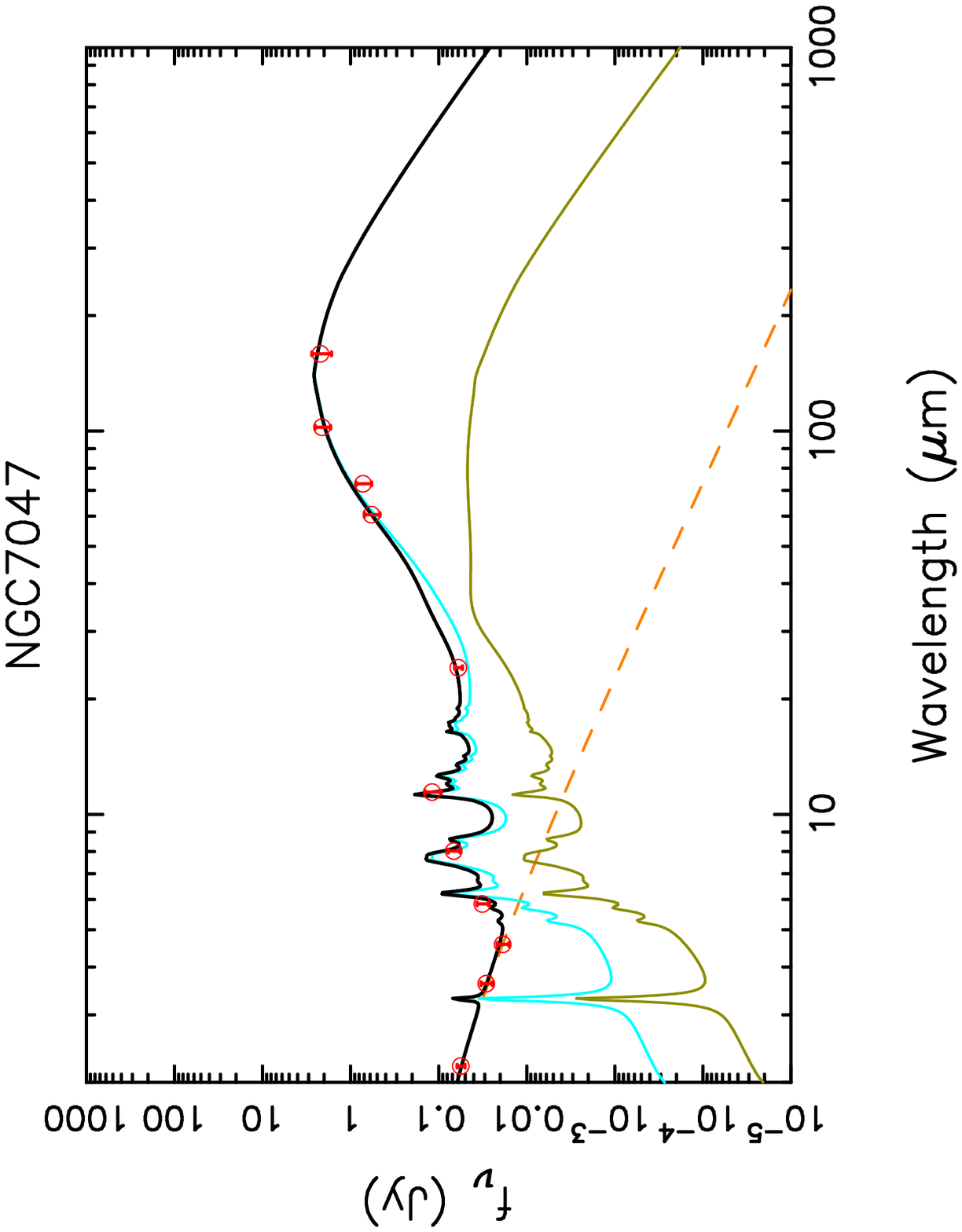}
\includegraphics{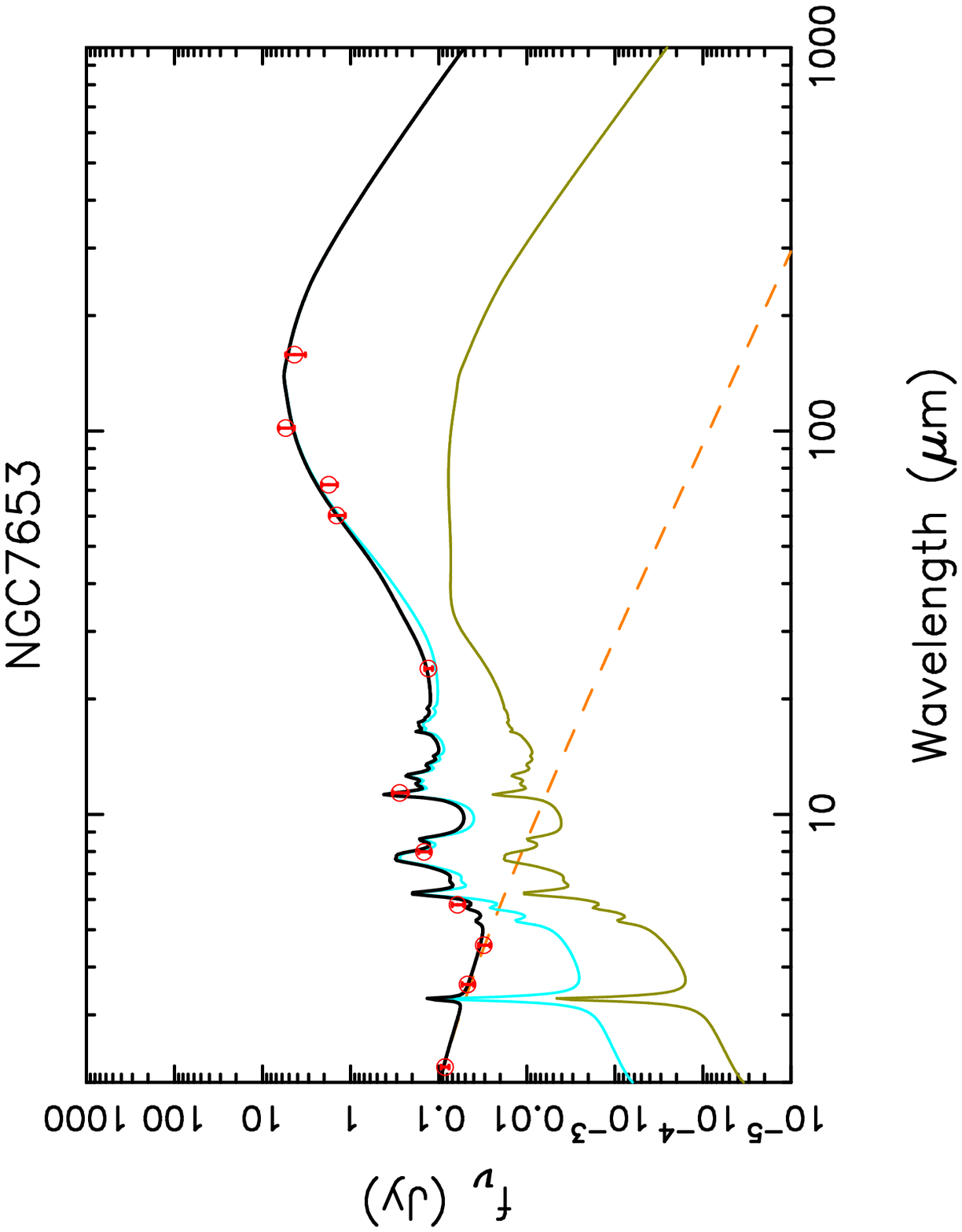}
\includegraphics{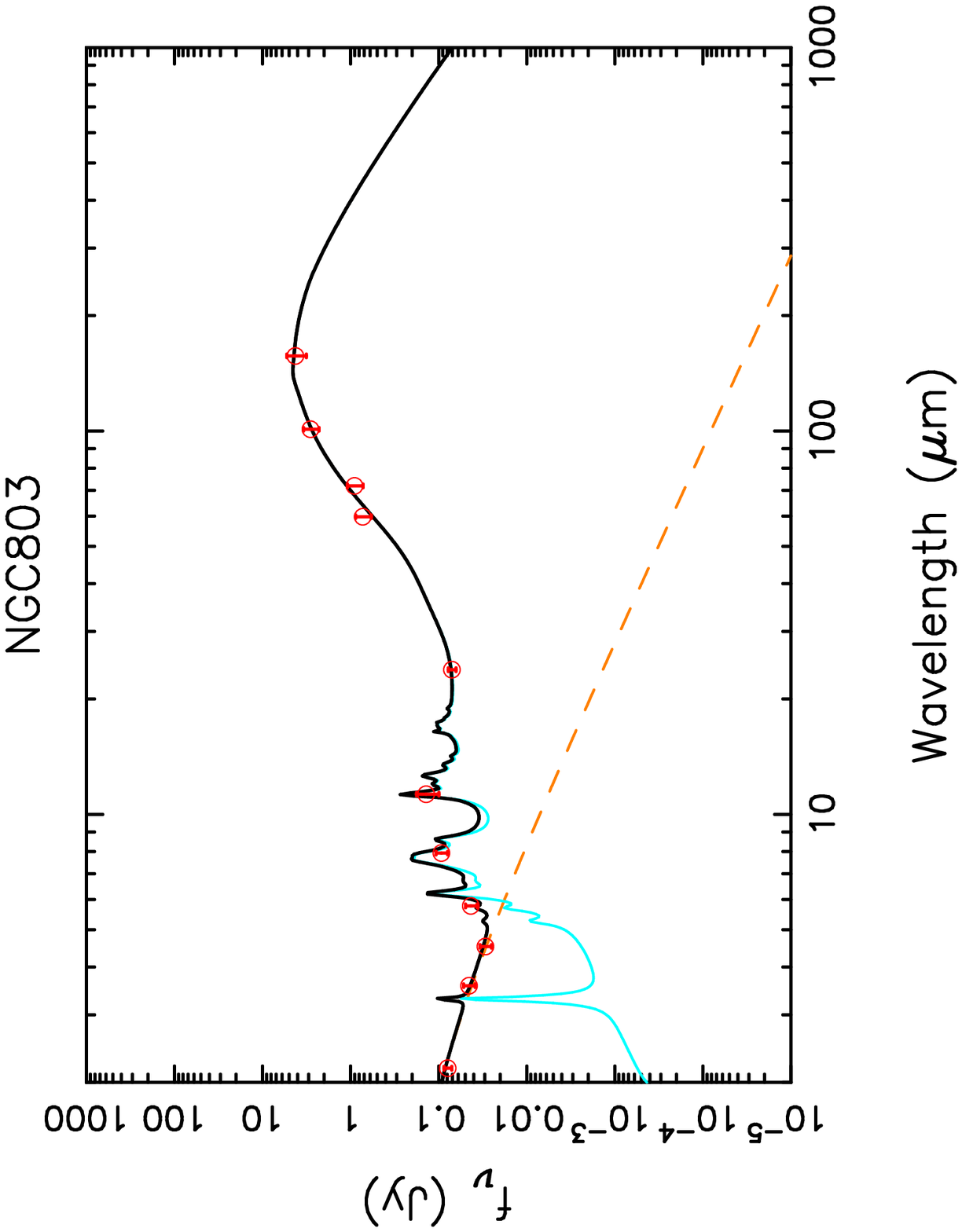}
\includegraphics{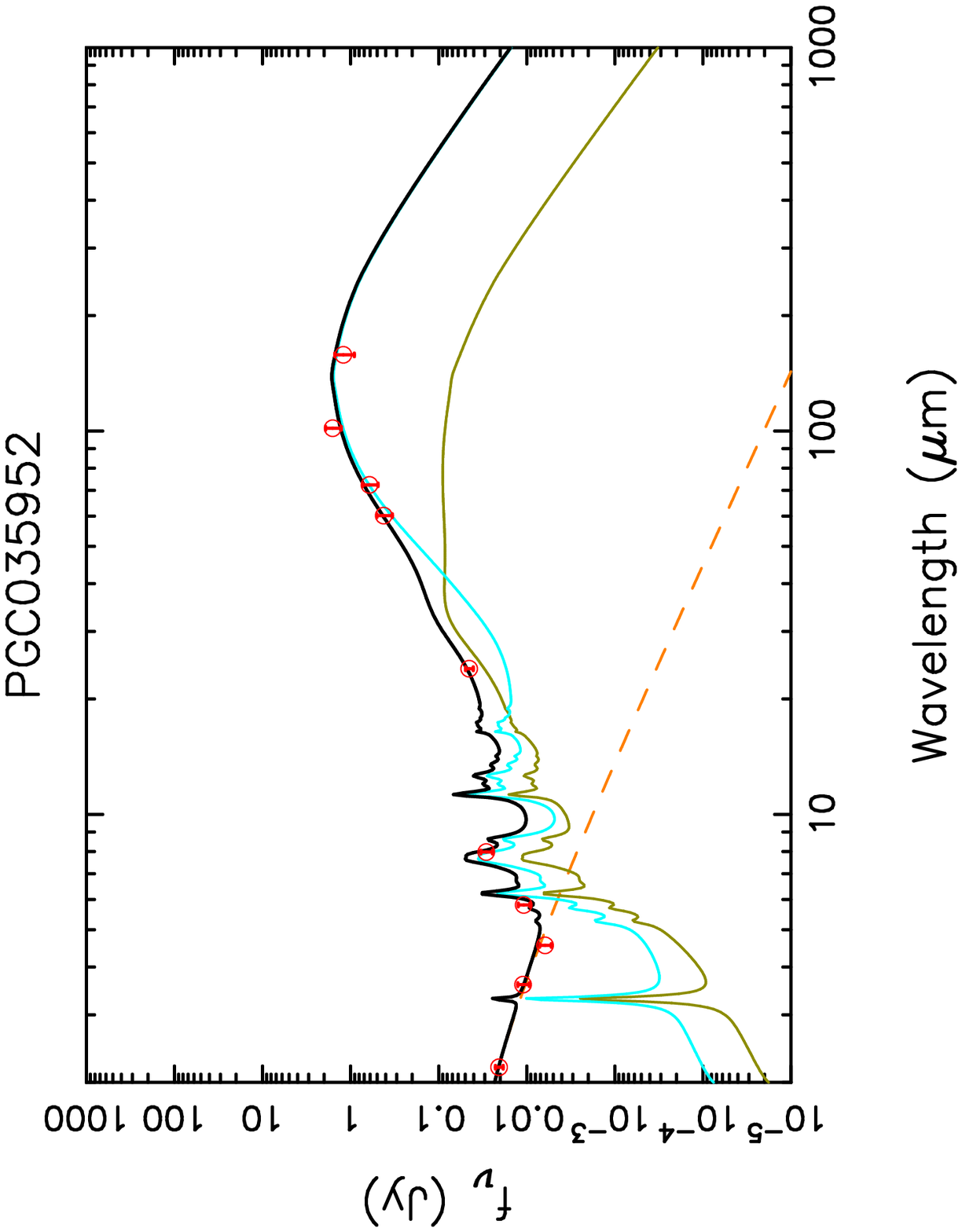}
\includegraphics{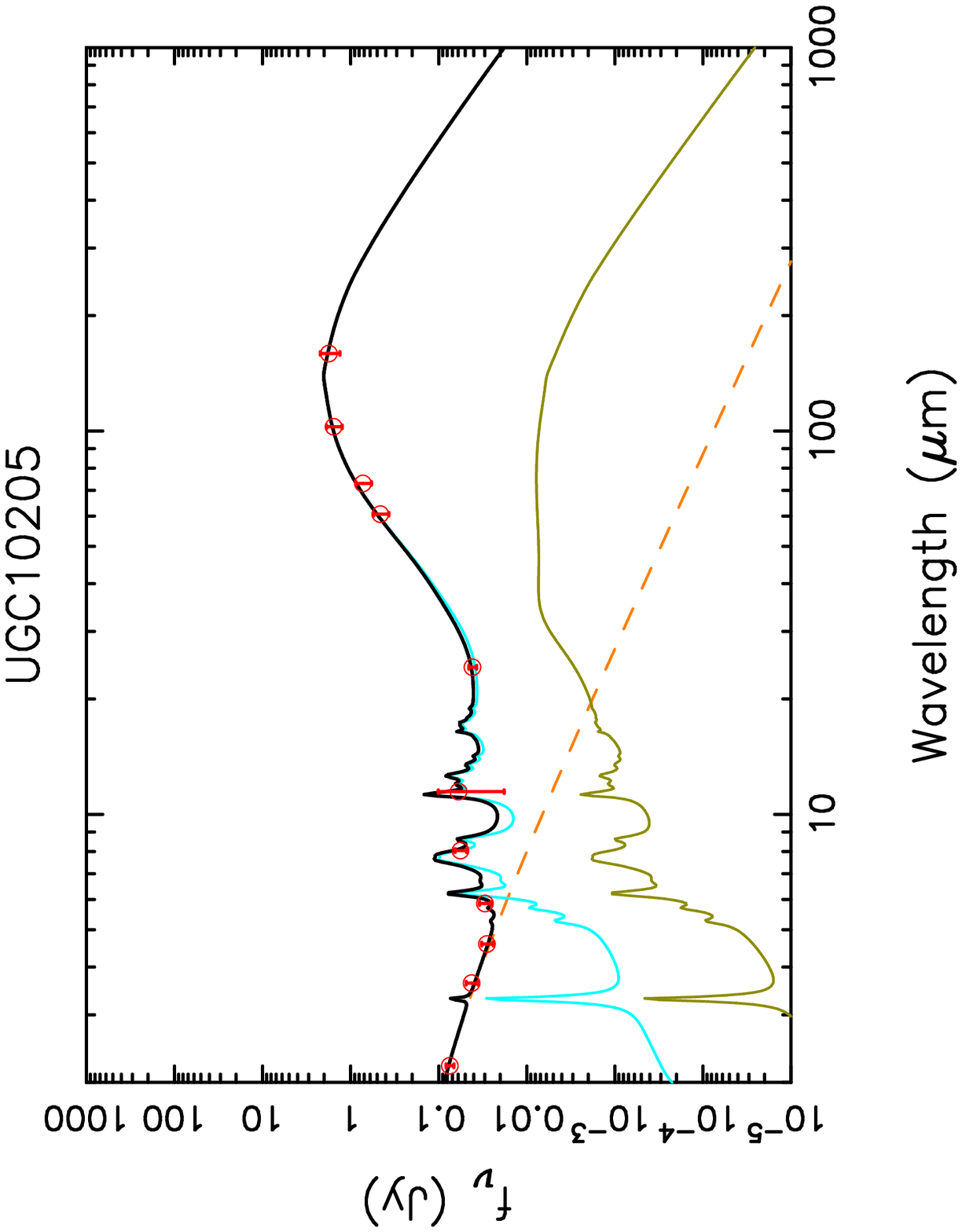}
\includegraphics{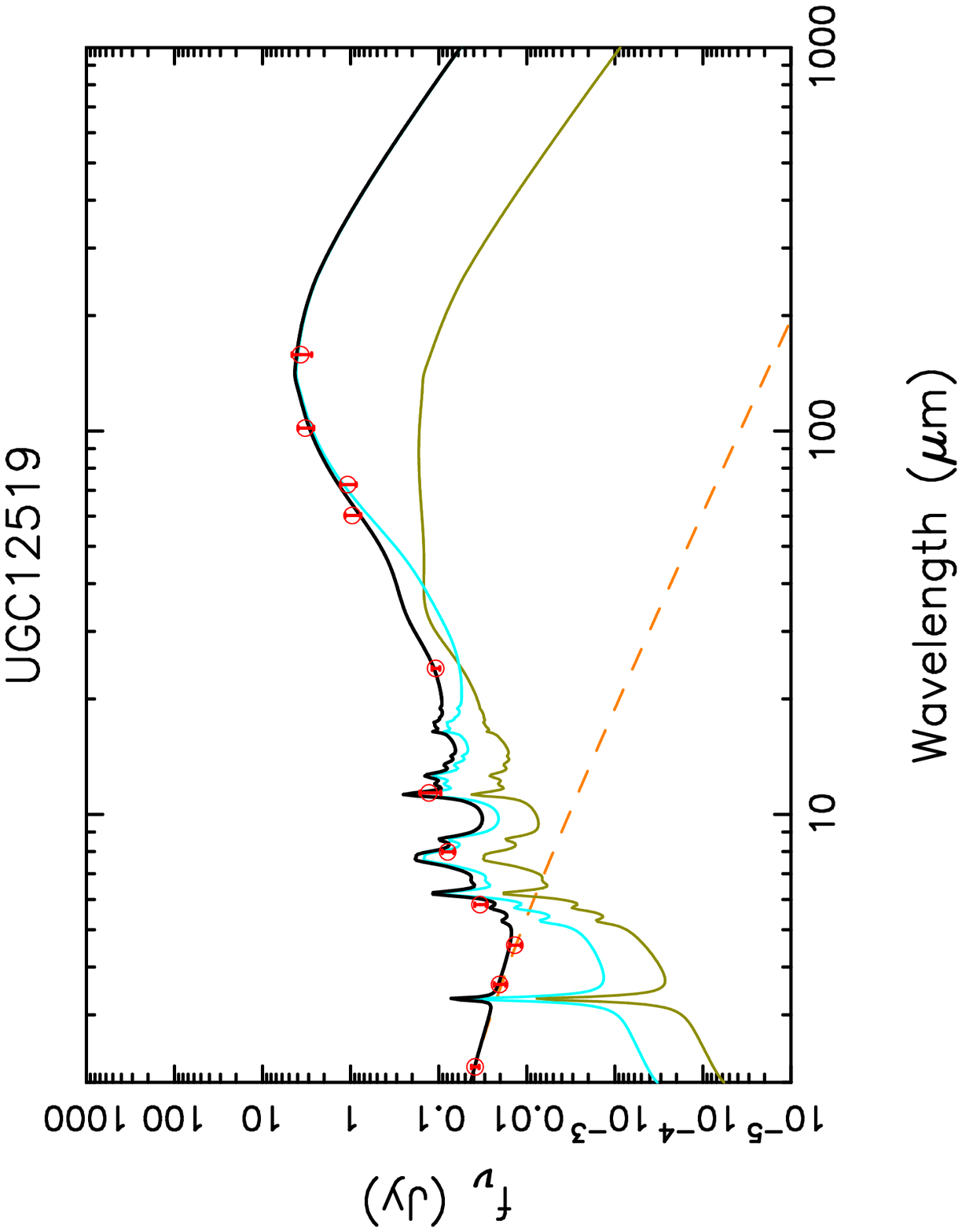}
\caption{Spectral energy distribution of SLUGS galaxies including data from
  2MASS, \iras, SCUBA,  IRAC and MIPS observations
  (open circles). The (estimated) error bars for most measurements
  are smaller than data points for most of the IRAC and MIPS
  data. Colors represent fits using the Draine \& Li (2007) models;
  the dashed orange line represents the estimated stellar flux, the green
  line the emission due to grains heated in photo-dissociation
  regions, cyan the emission from dust due to a single radiation
  field. The black line represents the sum of these components. The
  parameters for these fits, which also fit for $U_{max}$ the maximum
  starlight intensity, are presented in Table 6.
 }
\end{figure*}

\subsection {Comparison with SINGS}

\begin{figure}
\vspace{60mm}
\includegraphics{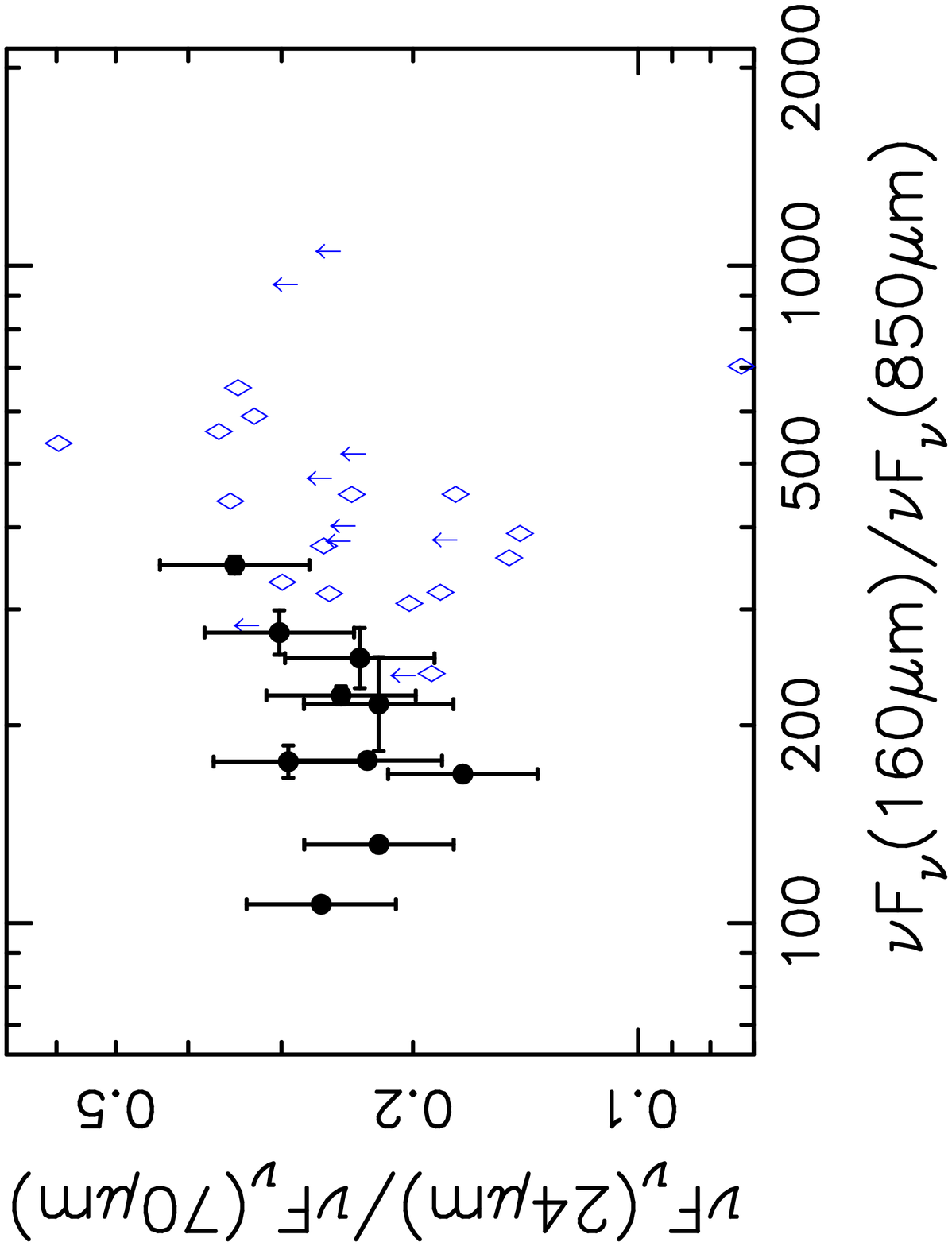}
\caption{Color-color distribution for galaxies with SCUBA measurements.
SLUGS galaxies are represented by black
solid circles while the diamonds denote SINGS galaxies with good SCUBA
measurements. The arrows represent upper limits of the 
$\nu$F${\nu}$(160$\mu$m)/$\nu$F${\nu}$(850$\mu$m) ratio of galaxies for
which the SCUBA measurements are probably underestimated (Draine et
al. 2007).
The segregation between both samples is very clear.
While both sample show a similar behavior in the 24/70
$\mu$m color which measures the mid-IR slope, the SLUGS galaxies show
flatter 160/850 $\mu$m slopes due to the larger contribution due to
cold dust.
}
\end{figure}

\begin{figure}
\vspace{70mm}
\includegraphics{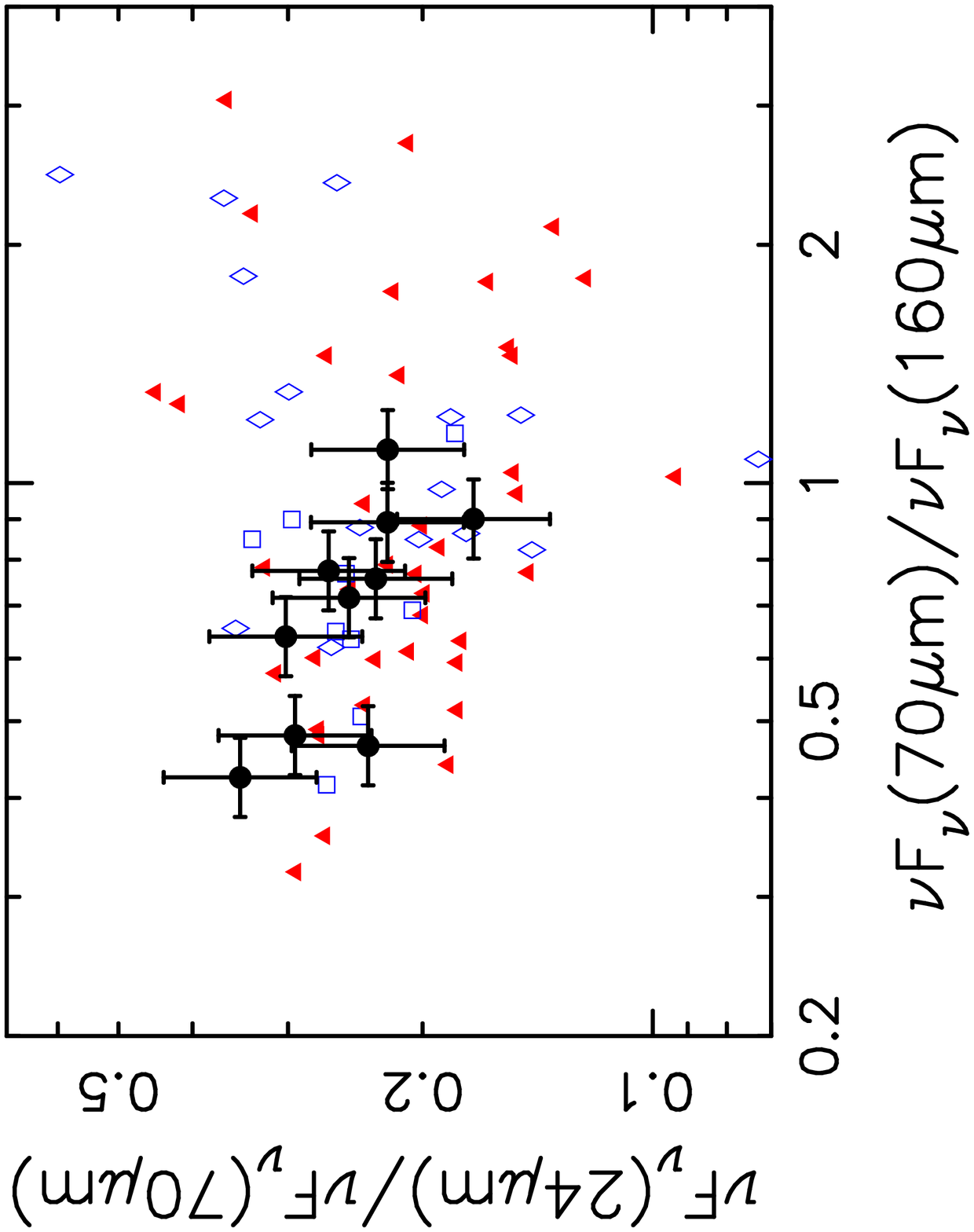}
\caption{Color-color distribution following Draine et al. (2007) of
  SLUGS galaxies (black circles) compared to SINGS galaxies with good SCUBA
  (diamonds) and lower confidence (suares) detections,  and without
  (triangles). The location of SLUGS galaxies towards lower values of
  ($\nu F_\nu)_{70\mu m}/(\nu F_\nu)_{160\mu m}$ is expected for
  galaxies with significant amounts of cold dust (Draine et al. 2007). 
}
\end{figure}

\subsubsection{Sample biases}
The combination of selection criteria (in addition to the small sample
sizes of SLUGS and SINGS), imposes a major limitation on the possible analyses
with these samples. As described by Vlahakis et
al. (2005), the parent sample for SLUGS is defined from the CfA1
Redshift Survey (Huchra et al. 1983), which in principle should be
unbiased as far as the cold-dust properties of galaxies are concerned
(cf.,Section 1).
The SLUGS galaxies selected for $Spitzer$ observations are a subset of
this larger sample.

The optical SLUGS sample was defined considering the number of
objects within a well defined region of the sky and with an
angular size that would fit inside the SCUBA field of view. The
parent sample contains a total of 97 galaxies, observations having been
taken for 81. Of the latter, 52 were detected in 850 $\mu$m, while 29 (36\%) only
provided upper limits. Early type (E+S0) galaxies comprise 14 (48 \%) of the
non-detections (corresponding to 17\% of the total observed sample),
spirals 8 (28\%) and the remainder are either multiple
systems or compact sources. The sample observed with $Spitzer$
contains 10 spiral galaxies and 1 elliptical, which is close to the
morphological mix of the SLUGS observed sample. Thus the small
proportion of early-types in the $Spitzer$ sample reflects in part the
low detection rate of these galaxies in the SLUGS sample (Vlahakis et
al. 2005).
There is a risk that the SLUGS/$Spitzer$ sample is not representative
of cold-dust properties of the local population of galaxies, being
biased towards galaxies with larger cold dust masses.

In a similar way, SINGS galaxies by design were not chosen to
represent a volume-limited sample of galaxies but rather to be
representative of the local morphology, luminosity and
$L_{FIR}/L_B$ parameter space (Kennicutt et al. 2003).  
Thus, when interpreting the comparisons that follow one must take into
account that neither sample is a ``fair'' representation of the local
mix of galaxy properties, so that the results may not be valid for the
general galaxy distribution. In particular, the observations of SINGS
galaxies with SCUBA did not follow any type of preselection based on
their dust properties. Of the 26 SINGS galaxies with SCUBA data, 13
were observed after the SINGS sample was defined, while data for 
the remainder were taken from the SCUBA archive by Dale et al. (2005).
However, we should note that Draine et al. (2007) identify 9 objects
which have lower confidence SCUBA observations which are
galaxies for which the data reduction removes extended emission (4
galaxies); galaxies for which SCUBA measurements do not cover the
total spatial extent where emission in 160 $\mu$m is detected (4
galaxies) and one galaxy for which the 850 $\mu$m emission is
dominated by an AGN. 

\subsubsection{Comparison between SINGS and SLUGS}

\begin{figure*}
\vspace{60mm}
\includegraphics{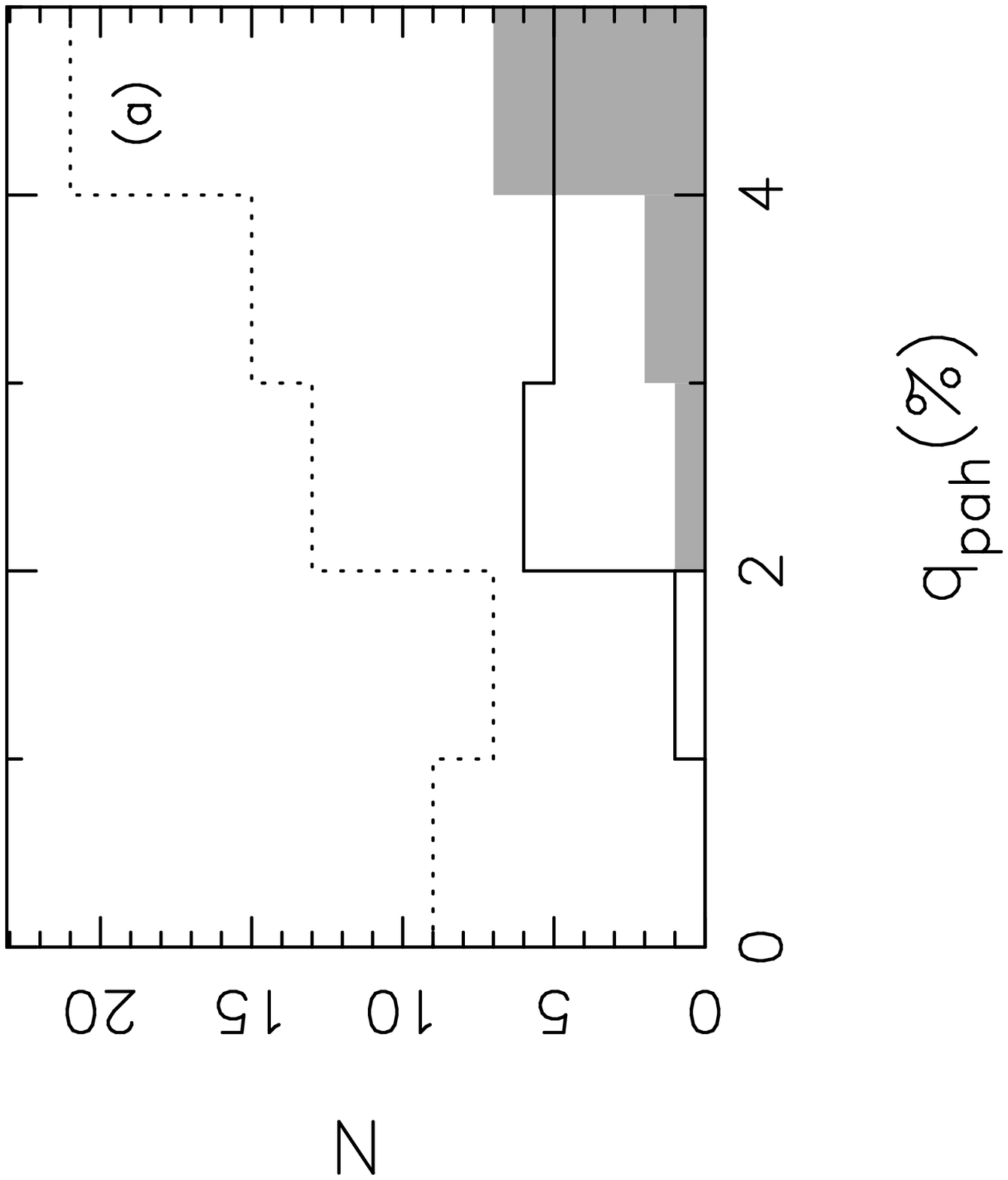}
\includegraphics{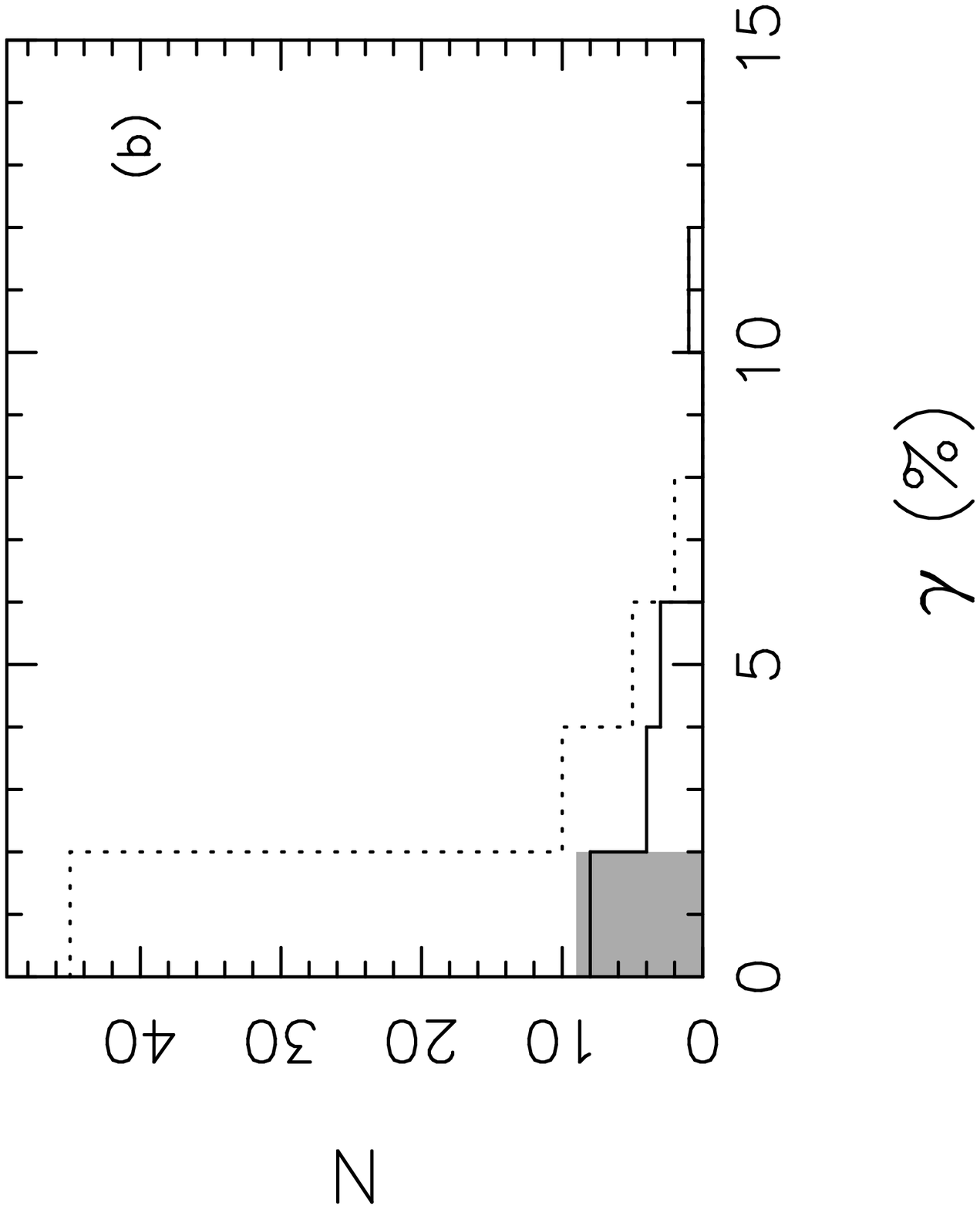}
\includegraphics{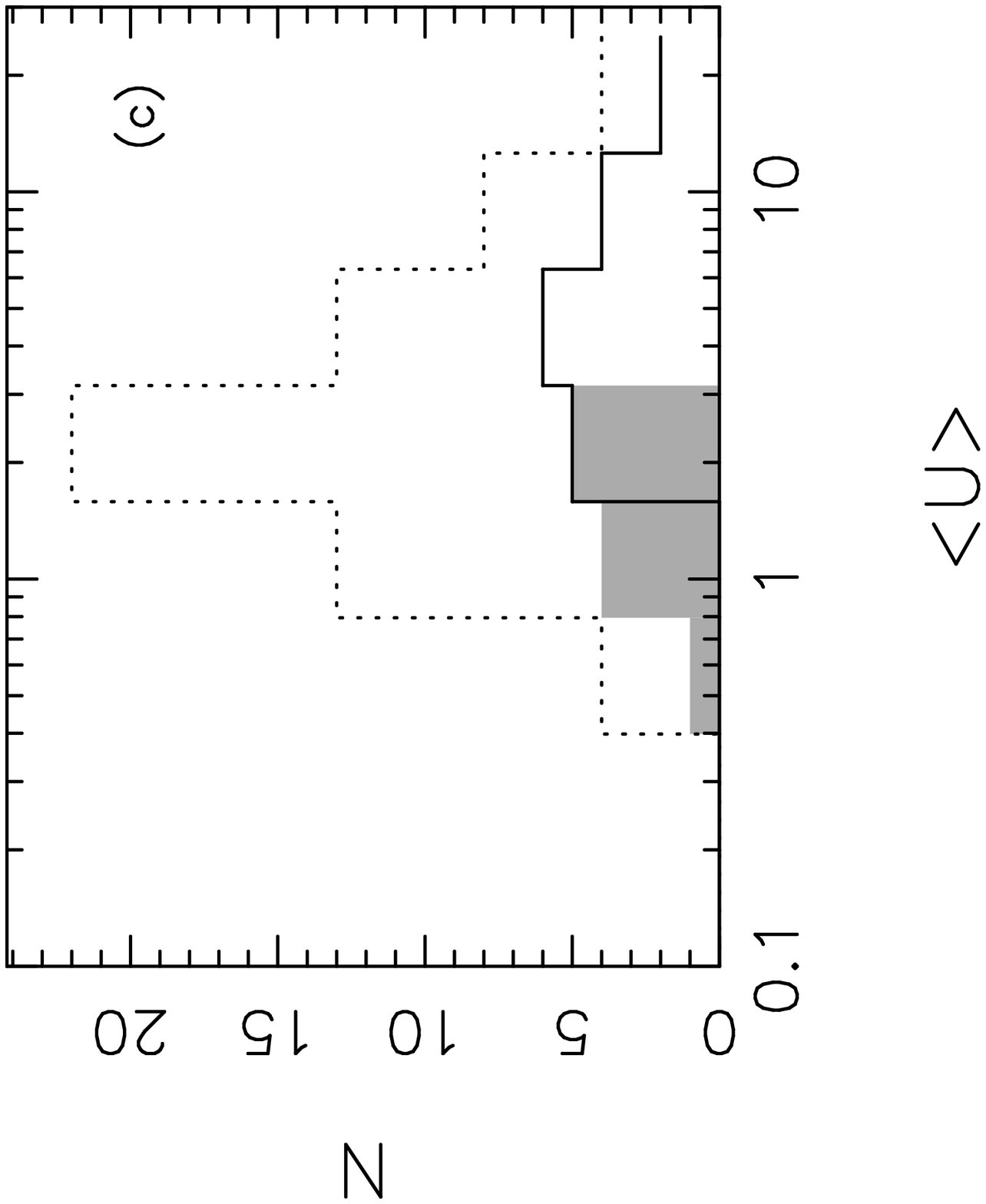}
\caption{Histograms comparing the parameters obtained by fitting the
  Draine et al. (2007) dust models to the SLUGS galaxies (solid  grey
  histograms), with results tabulated by Draine et al. (2007) for the
  full SINGS sample (dotted line histogram) and the subsample of 17 galaxies
  with SCUBA measurements (solid line histogram). This comparison uses
  the fit parameters obtained by fixing the value of $U_{max}$.
  Panel (a) shows the
  $q_{PAH}$ distribution, which measures the ratio between the dust mass
  in the form of aromatic grains and the total dust mass. Panel (b) shows
  the distribution of $\gamma$ 
  which quantifies the contribution towards the total dust emission
  from dust located in regions with intense starlight due to very
  hot stars. Panel (c) shows the average starlight radiation field
  ($<$U$>$) present in the galaxies' interstellar medium.
}
\end{figure*}
\begin{figure*}
\vspace{60mm}
\includegraphics{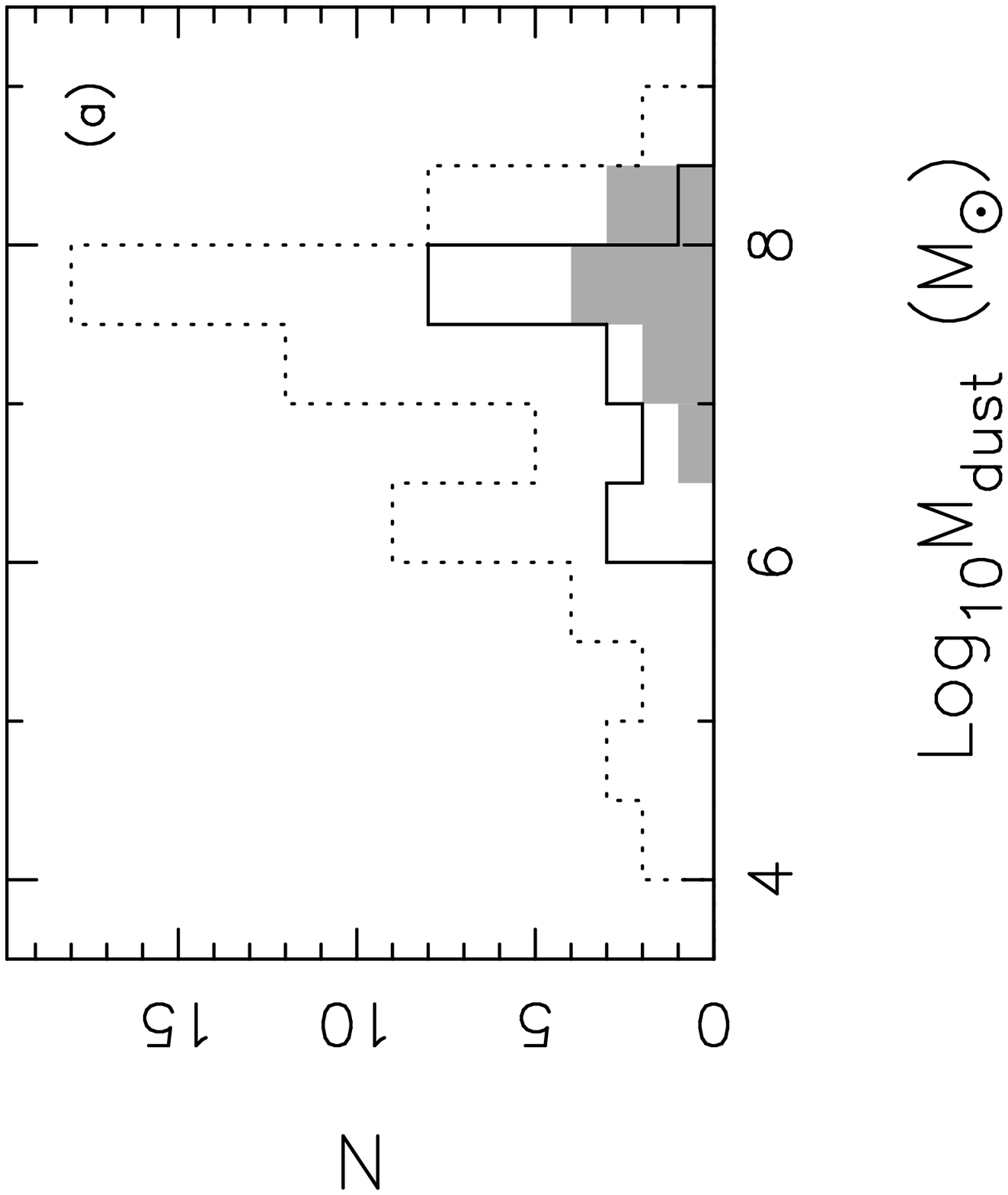}
\includegraphics{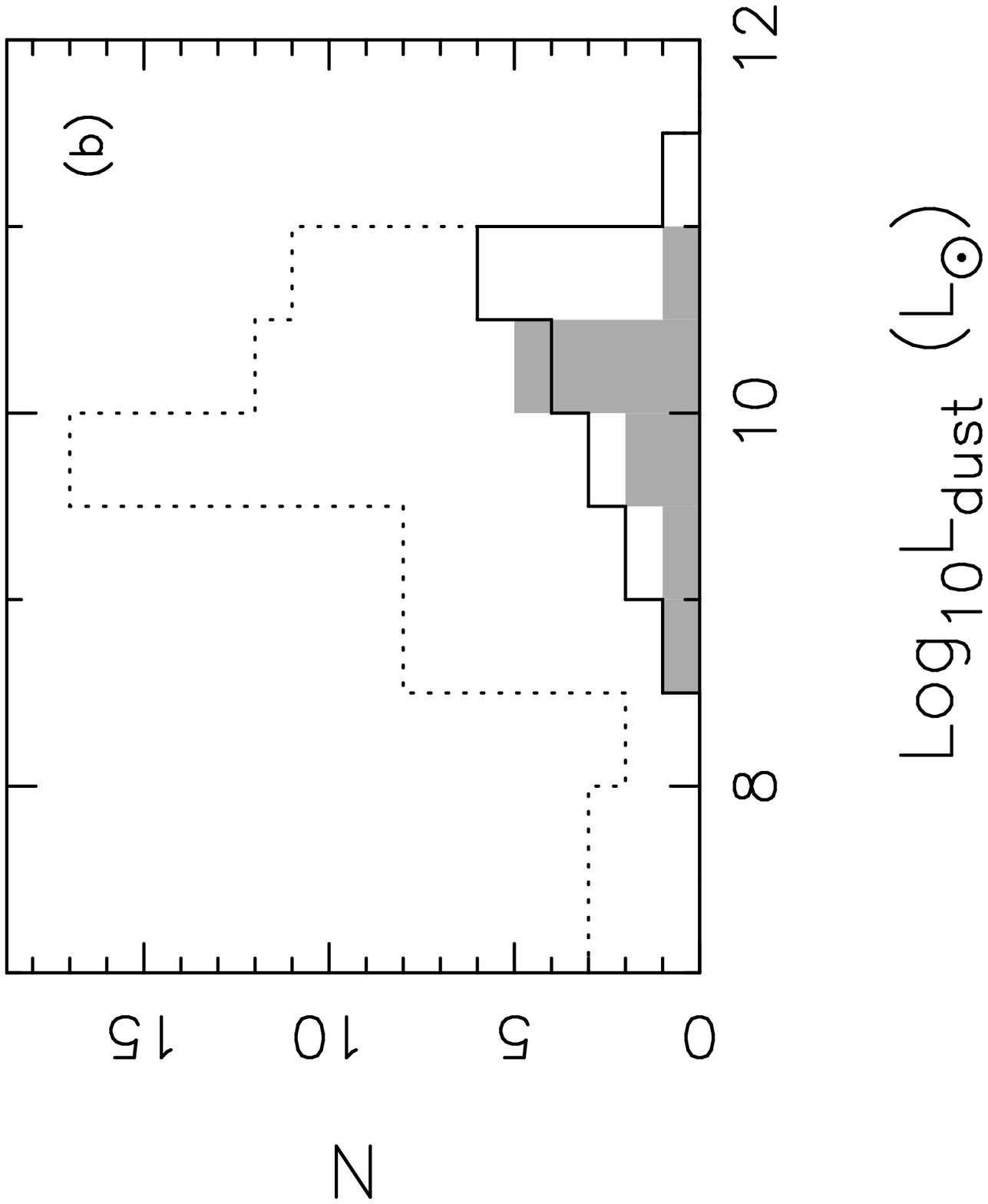}
\caption{Histograms comparing the estimated dust masses (left panel) and 
  luminosities (right panel) for SLUGS (grey histograms), 
  and SINGS galaxies (dotted line histogram). The 17 SINGS galaxies that have
  SCUBA measurements represented by the solid line histogram. The
  present sample has marginally higher dust masses than the SINGS
  sample, though the total dust luminosity is comparable.
}
\end{figure*}

One motivation for the new $Spitzer$ observations in this work was the
significantly different distribution observed using the sample of
SLUGS galaxies relative to SINGS when considering the \iras ~25 to 60
$\mu$m and 100 to 850 $\mu$m color-color distribution, which
measures the relative intensity of warm to cold dust emission.
Figure 5 shows the analogous distribution using MIPS data for both
samples in the 24 to 70 $\mu$m and 160 to 850 $\mu$m colors, where
it is clearly seen that SLUGS (solid circles) and SINGS galaxies
{\it{with}} SCUBA measurements
(diamonds) occupy mainly different regions (we should note
that the parent SLUGS sample covers the entire \iras ~25 to 60 $\mu$m color
range). Even when including SINGS galaxies for which the 850 $\mu$m
may be underestimated, thus providing upper limits to the 
$\nu$F$_\nu$(160$\mu$m)/$\nu$F$_\nu$(850$\mu$m) ratio
(noted by the arrows), the locus of SLUGS galaxies still differs from
that of the increased sample.

We now consider several combinations of measureable parameters to
examine whether this difference found between SLUGS and the SINGS
subsample with SCUBA data also extends to full SINGS sample when
galaxies with no SCUBA data are included.
The color-color plot using MIPS observations is presented in Figure 6,
again showing SLUGS and SINGS galaxies. As in Fig. 5, we
discriminate galaxies with good SCUBA observations (diamonds) from
those with lower-confidence level measurements (squares) and no SCUBA
observations (triangles).  
The plot shows that SLUGS galaxies are found towards lower values of 
$\nu$F$_\nu$(70$\mu$m)/$\nu$F$_\nu$(160$\mu$m) which is the expected 
behavior for galaxies with cold dust, though not reaching the levels
of NGC~4725 and NGC~2841, which are the most extreme cases in the
SINGS sample. 
On the other hand, the distribution of the 
$\nu$F$_\nu$(24$\mu$m)/$\nu$F$_\nu$(70$\mu$m)
colors of the SLUGS sample
is similar to that of the SINGS sample.

The data in Table 7 are used to compare the properties of SLUGS
galaxies with the Draine et al. (2007) fits for SINGS where the
maximum intensity of the starlight ($U_{max}$) is fixed to 10$^{6}$.
The SED fits calculated in Section 3.2 suggest that most galaxies in
the SLUGS sample have PAH emission, which is further corroborated by
the analysis of Bendo et al. (2008). The frequency distribution of the
amount of dust contained in small aromatic grains ($q_{PAH}$) is 
shown in Figure 7(a). The SLUGS sample
is represented by the solid grey histogram while Draine et
al. (2007) measurements for SINGS galaxies are outlined
as the dotted histogram for the entire sample and the solid line
histogram for the SCUBA sub-sample
of 17 galaxies. The distribution of aromatic grain indices for SLUGS
and the full SINGS 
is similar in spite of the small number of SLUGS objects.
Figure 7(b) shows the distribution of $\gamma$, which 
quantifies the contribution due to starlight
re-radiated in photo-dissociation regions relative to the idealized
single and constant radiation field $U_{min}$. The  $\gamma$
distribution for SLUGS has a median value of $\sim$ 0.6\%, and the
galaxies are all concentrated in a single bin in the histogram; SINGS
galaxies have $\gamma$ $\sim$ 1\%. This low value of $\gamma$ for
SLUGS has direct   
implications for the estimate of the average starlight intensity shown
in Figure 7(c) which is lower ($<U>_{slugs}\sim$ 1.3) than what is
measured for SINGS ($<U>_{sings} \sim$ 2.5).

Figure 8(a) shows the distribution of total dust masses. SLUGS
galaxies have masses ranging from 
$\sim$ 10$^{6.5}~M_\odot$ to  10$^{8.5}~M_\odot$, with a median of
10$^{7.9}~M_\odot$,  $\sim$ 3  $\times$ higher than is measured
for the entire sample of SINGS galaxies (10$^{7.4}~M_\odot$).
However,  as noted by Draine et al. (2007), for galaxies without
sub-mm data the fitting procedure has to be more constrained otherwise
the resulting fits may produce $larger$ cold dust masses. However,
Draine et al. (2007) find that the mass estimates for $\sim$ 65\% of
their sample agree within a factor of 1.5 when fits using and
discarding the sub-mm data are compared. Given the uncertainties,
the observed difference between SLUGS and SINGS is probably not
significant, corroborated by the similar masses of SINGS galaxies
with sub-mm data (10$^{7.6}~M_\odot$).
Figure 8(b) shows the total dust luminosities, which for SLUGS galaxies
($log_{10} L/L_\odot \sim 10.2$) are 
somewhat higher than for the entire SINGS sample ($log_{10} L/ L_\odot
\sim 9.7$) (though the distribution is virtually identical to that of
the  SINGS SCUBA subsample $log_{10} L/ L_\odot \sim 10.3$).  
Differences between the distribution of the dust model parameters were
examined using the two-sided Kolmolgorov-Smirnov test (e.g., Press et
al. 2002), and the only significant difference between both samples
is found for the average starlight field $<U>$ where the probability
of both distributions being drawn from the same parent sample is
$\lesssim$ 10$^{-5}$.  

On average, the SLUGS
galaxies have a warm dust component with a median temperature of
$T_w$=54.7$\pm$1.7~K and a colder one with $T_c$= 18.5$\pm$1.1~K.
Both estimates agree within 2 $\sigma$ with  the results of Vlahakis et
al. (2005) who find $T_w$=47.4$\pm$2.4~K and $T_c$= 20.2$\pm$0.5~K
respectively.
The warm dust temperature is comparable to that
measured by Hinz et al. (2007) for UGC~6879,
while the the average value we find for the cold dust component
is comparable to the Hinz et al. (2006)
measurement for the starbursting dwarf galaxy UGC~10445.
Similarly to the findings of Vlahakis et al. (2005) and Hinz et
al. (2007), most of
the dust in these galaxies is cold, with  $M_{cold}/M_{warm}$
$\sim$ 1156$\pm$658.  The distribution of the ratio between the cold
and warm dust components for SLUGS and SINGS is shown in Figure 9.
Although there is some overlap in the mass ratios, SINGS galaxies with
SCUBA measurements tend to have a smaller contribution due to cold
dust with a ratio of  538$\pm$305.

The ratio between the total emission integrated from 5--1000 $\mu$m,
which includes the important contribution due to aromatic emission
(e.g., Rieke et al. 2009) and the infrared luminosity in the range
100-1000 $\mu$m, which is mostly dominated by the cold dust component,
is shown in Fig. 10. SLUGS galaxies are 
represented by the solid black circles, SINGS galaxies with SCUBA
measurements by diamonds or open squares (in the case of low
significance level measurements); the triangles
represent SINGS galaxies for which the far-IR is not as well
constrained because of the absence of SCUBA data. 
Also shown are the sample of local LIRGs and ULIRGs (squares) and
templates for 10$^{9.75}$ $\leq L \leq$ 10$^{11}$ $L_\odot$ of
Rieke et al. (2009) represented by stars.
The SLUGS galaxies populate a region of lower luminosities and with a
higher proportion of far-IR emission than the sample of SINGS
galaxies with SCUBA measurements. 
However, it is interesting to note
that the constrained Draine \& Li (2007) models place many of
the SINGS galaxies without SCUBA measurements in roughly the same
parameter space as the SLUGS galaxies. The combined SLUGS and SINGS
and LIRG/ULIRG samples show a trend where the amount of energy in the
far-IR becomes more important as lower luminosities are reached. This
trend is also seen in the Rieke et al. (2009) templates.

\begin{figure}
\vspace{60mm}
\includegraphics{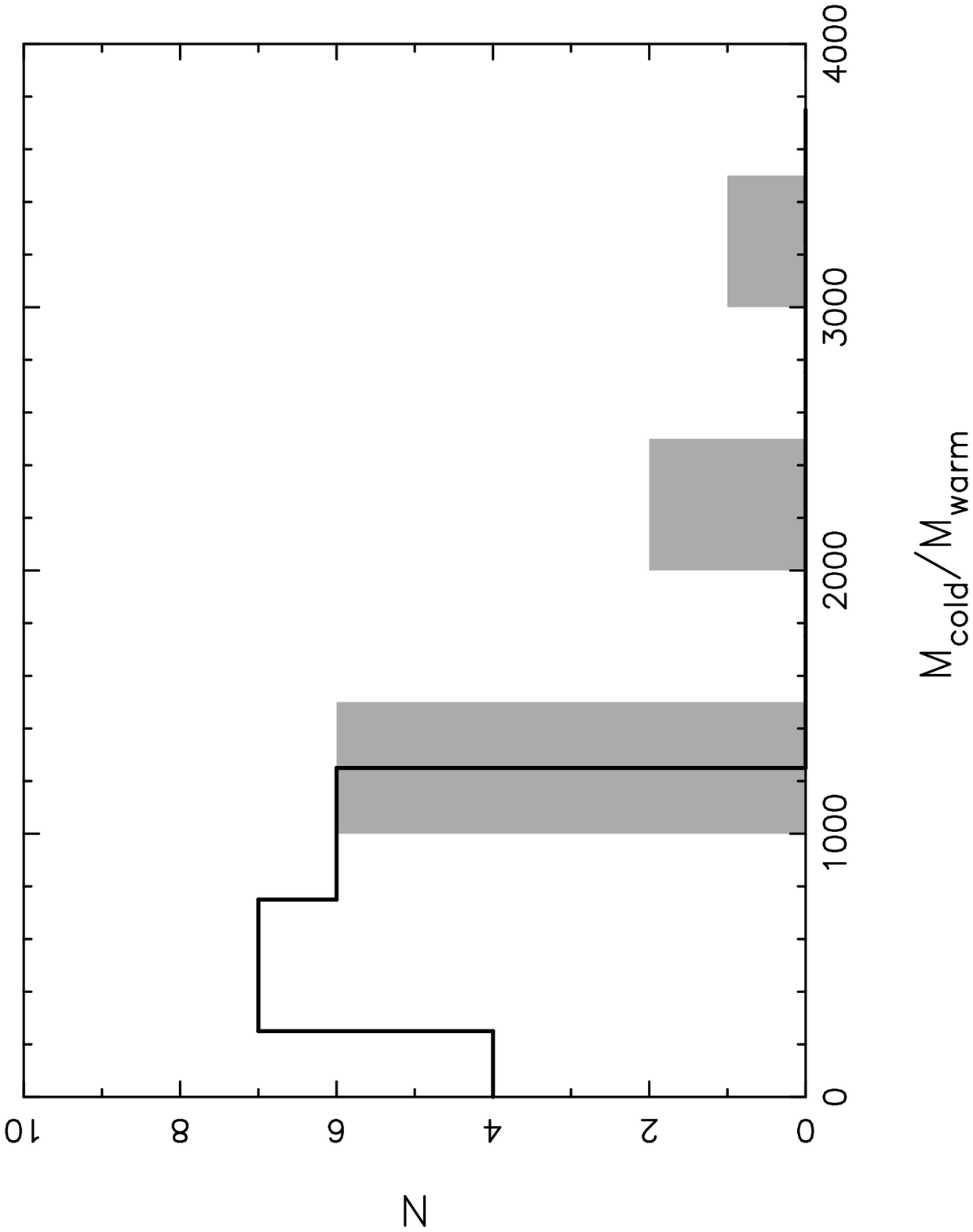}
\caption{Distribution of the mass ratio between cold (T$\sim$19K) and
  warm (T $\sim$ 55 K) dust estimated from modified blackbody
  fits for SINGS and SLUGS galaxies with SCUBA data. The filled
  histogram represents SLUGS galaxies and the open histogram SINGS
  galaxies. Although there is significant overlap, SINGS galaxies tend
  to have smaller mass ratios compared to SLUGS.
}
\end{figure}

\begin{figure}
\vspace{60mm}
\includegraphics{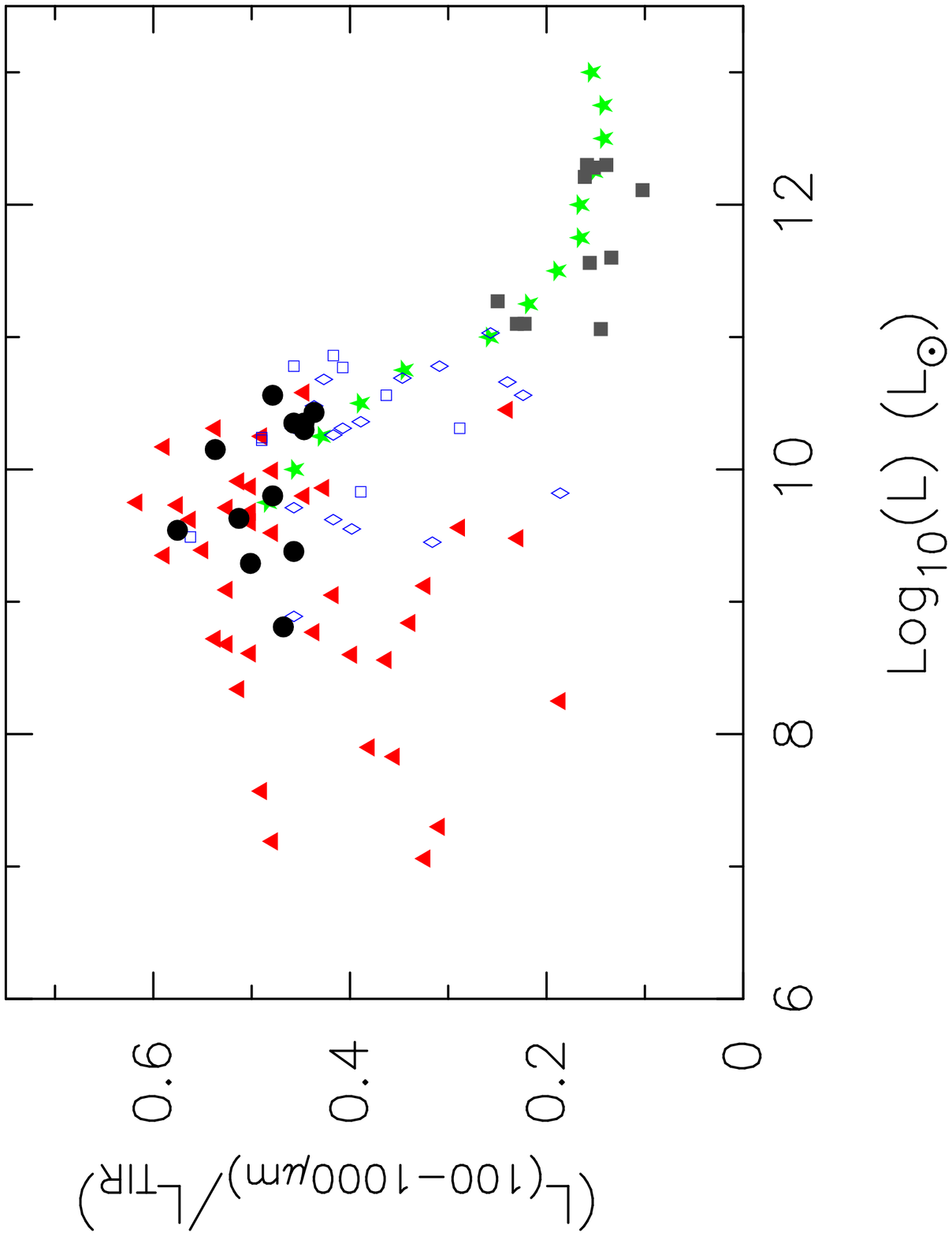}
\caption{Ratio between the luminosity emitted in the far-IR (100-1000 $\mu$m)
  and the total luminosity (5-1000 $\mu$m) for SLUGS galaxies (black circles),
  SINGS galaxies with SCUBA detections (diamonds), those without
  (triangles); the Rieke et al. (2008) sample of local LIRGs and
  ULIRGs (solid squares) and the Rieke et al. (2008) templates
  (stars) covering a range of luminosities from 10$^{9.5}$ to
  10$^{13}$ $L_\odot$. The local LIRGs and ULIRGs extend to lower
  values the cold/total emission measured by SINGS and SLUGS. The
  behavior of the Rieke et al. (2008) templates show that the three
  samples are complementary. These also suggest that galaxies with
  lower IR luminosities tend to have a larger contribution of their
  emission in the far-IR. 
}
\end{figure}

The new observations reported in this paper suggest that there is a wider
variety of sub-mm properties than encompassed by SINGS alone, with a
larger fraction of galaxies in the nearby universe 
presenting significant amounts of dust at temperatures $\sim$ 18
K. The luminosities of these galaxies  (Log($L/L_\odot$) $\sim$ 10.2)
are comparable to some of the least luminous objects detected in 
the SCUBA Half Degree Extragalactic Survey (SHADES), a blank field  
survey which scanned several patches of sky (Mortier et al. 2005), in
contrast to SLUGS, which targeted optically-detected galaxies.
The lower luminosity galaxies in SHADES are characterized as having
``cirrus'' SEDs (e.g., Clements et al. 2008) and make up $\sim$ 25 \% of
the sample of sub-mm galaxies (Clements et al. 2005), suggesting that
the local samples are undersampling this type of galaxy.

\section{Conclusion}

We trace the Spectral Energy Distribution for 11 galaxies providing
an unbiased sample of sub-mm properties (from the SLUGS selection). We
combine new IRAC
and MIPS data obtained with the {\it{Spitzer Space Telescope}} with
existing 850 $\mu$m, \iras~ and near-IR measurements. 
The new measurements confirm the presence of cold (T$\sim$18K) dust in 10 of
these galaxies.

Using fits of Draine \& Li (2007) models we show
that the contribution due to light reprocessed by
photo-dissociation regions is not as important in our sample as it is
for SINGS galaxies. 
The models also suggest that the total IR emission of SLUGS
galaxies has a small contribution due to PAH lines emitted in
photo-dissociation regions. 
However the proportion of the
total light contributed by the mid-IR emission is smaller than that
measured for the SINGS sample. 
The typical total dust masses obtained
using Draine \& Li (2007) models are of 10$^{7.9} M_\odot$ and dust
luminosities are $\sim$ 10$^{10} L_\odot$. 
The most significant difference between the
samples is the lower average intensity of the general radiation
field of SLUGS compared to SINGS galaxies.
Although the uncertainties are large, SLUGS galaxies show $\sim$ 2
$\times$ larger proportion of cold dust relative to warm than do the
SINGS galaxies. 

The present data combined with previous observations show
that cold dust is present in all types of spirals being the major
contributor to their far-infrared luminosity (Tuffs and Popescu
2003). Although the 160 $\mu$m data are somewhat shallow, they suggest
that qualitatively the scale length distribution of cold dust is at least 
comparable to what is measured in the mid-infrared and in the visible
(e.g., Alton et al. 1998; Davies
et al. 1999; Tuffs et al. 2002a, 2002b; Hinz et al. 2006). 
Since the amount of dust in a galaxy measures the quantity of heavy  
elements in the ISM, it gives a complementary measure of the average
metallicity of galaxy (Dunne et al. 2000). Thus the presence of cold
dust suggests that the extended HI in these galaxies is generally
not metal-poor. 

\acknowledgements
C.N.A.W. thanks Z. Balog, M. Blaylock and M. Block for their guidance in the
reduction of IRAC and MIPS data, T. Lauer for tips on
photometry and J. Stansberry for discussions on the calibration of
$Spitzer$ images. ELF acknowledges the support from 
the Spitzer Space Telescope Fellowship Program through a contract issued
by the Jet Propulsion Laboratory, California Institute of
TechnologyJPL/Caltech and National Aeronautics and Space Administration. 
This work is based in part on observations made with the 
{\it{Spitzer Space Telescope}}, which is operated by the JPL/Caltech under
NASA contract 1407. Support for this work was provided by NASA through
JPL/Caltech contract 1255094 and by MAG5-12318 from NASA/Goddard to
the University of Arizona.
This research has made use of NASA's Astrophysics Data System (ADS) and
the NASA/Infrared Processing and Analysis Center Extragalactic
Database (NED) which is operated by the JPL/Caltech, under
contract with NASA. Data also come from the Two Micron All Sky Survey,
a joint project of the University of Massachusetts and IPAC/Caltech,
funded by NASA and the NSF. 
We acknowledge the usage of the HyperLeda database (http://leda.univ-lyon1.fr).


\begin{references}
\normalsize
\reference {} Bell, E. F. 2003, ApJ, 586, 794
\reference {} Bell, E. F., McIntosh, D., Katz, N., \& Weinberg,
M. D. 2003, ApJS, 149, 289
\reference {} Bendo, G. J., et al. 2008, MNRAS, in press, (arXiv 0806.2758)
\reference {} Bertin, E, \& Arnouts, S. 1996, A\&AS, 117, 393 
\reference {} Blain, A. W., Chapman, S. C., Smail, I.m \& Ivison,
R. 2004, ApJ, 611, 52
\reference {} Busko, I. 1996, in ``Astronomical Data Analysis Software
and Systems V'', eds. G. H. Jacoby \& J. Barnes, ASP Conference
Series, Vol. 101, 139
\reference {} Chary, R. R., \& Elbaz, D. 2001, ApJ, 556, 562
\reference {} Clements, D. L., Farrah, D/, Rowan-Robinson, M., Afonso,
J., Priddey, R., \& Fox, M. 2005, MNRAS, 363, 229
\reference {} Clements, D. L., et al. 2008, MNRAS, 387, 247
\reference {} Cutri, R. et al. 2006, ``Explanatory Supplement to the
2MASS all Sky Data Release and Extended Mission Products'',
{\it{http://www.ipac.caltech.edu/2mass/releases/allsky/\\doc/explsup.html}},
Section IV.2
\reference {} Dale, D. A., Helou, G., Contursi, A., Silbermann, N. A.,
Kolhatkar, S. 2001, ApJ, 549, 215
\reference {} Dale, D. A. \& Helou, G. 2002, ApJ, 576, 159
\reference {} Dale, D. A., et al.  2005, ApJ, 633, 857
\reference {} Dale, D. A., et al.  2007, ApJ, 655, 863
\reference {} D\'esert, F. X., Boulanger, F., \& Puget, J.-L. 1990,
A\&A, 237, 215
\reference {} Devereux, N. A., \& Young, J. S., 1990, ApJ, 359, 42
\reference {} Draine, B. T., \& Li, A.  2001, ApJ, 551, 807
\reference {} Draine, B. T., \& Li, A.  2007, ApJ, 657, 810
\reference {} Draine, B. T., et al. 2007, ApJ, 663, 866
\reference {} Dunne, L., Eales, S. Edmunds, M, Ivison, R., Alexander,
P, \& Clemens, D. L. 2000, MNRAS, 315, 115
\reference {} Elbaz, D., Cezarsky, C. J., Chanial, P., Aussel, H.,
Franceschini, A., Fadda, D., \& Chary, R. R. A\&A 384, 848
\reference {} Engelbracht, C. W., et al. 2007, PASP, 119, 994
\reference {} Fazio, G. G., et al. 2004, ApJS, 154, 10
\reference {} Finkbeiner, D. P., Davis, M., \& Schlegel, D. J. 1999,
ApJ, 524, 867 
\reference {} Galliano, F., Madden, S. C., Jones, A. P., Wilson,
C. D., \&  Bernard, J. -P. 2005, A\&A, 434, 867
\reference {} Giovanini, G., Taylor, G. G., Feretti, L., Cotton,
W. D., Lara, L. \& Venturi, T. 2005, ApJ, 618, 635
\reference {} Gordon, K. D., et al. 2005, PASP, 117, 503
\reference {} Gordon, K. D., et al. 2007, PASP, 119, 1019
\reference {} Guiderdoni, B., Hivon, E., Bouchet, F. R., \& Maffei, B.
1998, MNRAS, 295, 877
Fronti\'eres), p. 521
\reference {} Haas, M., Klass, U., \& Bianchi, S. 2002, A\&A, 385,L23
\reference {} Haas, M., Lemke, D., Stickel, M. Hippelein, H., Kunkel,
M., Herbstmeier, U., \& Mattila, K. 1998, A\&A, 338, 33
\reference {} Helou, G. 1986, ApJ, 311, L33
\reference {} Hippelein, H., Haas, M., Tuffs, R. J., Lemke, D.,
Stickel, M., Klass, U. \& V\"olk, H. J. 2003, A\&A, 407, 137
\reference {} Hinz, J. L., Misselt, K., Rieke, G. H., Rieke, M. J.,
Smith, P. S., Gordon, K. D.  2006, ApJ, 651, 874
\reference {} Hinz, J. L., Rieke, M. J., Rieke, G. H., Willmer, C. N. A.,
Misselt, K. Engelbracht, C. W., Blaylock, M. \& Pickering, T. E. 2007,
ApJ, 663, 895
\reference {} Holland, W. S., et al. 1999, MNRAS, 303, 659
\reference {} Huchra, J. P., Davis, M., Latham, D. W., \& Tonry,
J. 1983, ApJS, 52, 89
\reference {} James, A., Dunne, L., Eales, S., \& Edmunds, M. G. 2002,
MNRAS, 335, 753
\reference {} Jarrett, T. H., Chester, T., Cutri, R., Schneider, S.,
Skrutskie, M., \& Huchra, J. P.  2000, AJ, 119, 2498
\reference {} Jedrzejewski, R. I. 1987, MNRAS, 226, 747
\reference {} Kessler, M.F., et al. 1996, A\&A, 315, L27
\reference {} Kennicutt, R. C., et al.  2003, PASP, 115, 928
\reference {} Lagache, G., Dole, H., \& Puget, J.-L. 2003, MNRAS, 338, 555
\reference {} Lagache, G., et al.  2004, ApJS, 154, 112
\reference {} Lagache, G., Puget, J.-L., Dole, H. 2005, ARAA, 43, 727
\reference {} Laurent, O., Mirabel, I. F., Charmandaris, V., Gallais,
P., Madden, S. C., Sauvage, M., Vigroux, L., \& Cezarsky, C. 2000a,
A\&A, 359, 887
\reference {} Laurent, O., Mirabel, I. F., Charmandaris, V., Le
Floc'h, E., Lutz, D., \& Genzel, R. 2000b, in ``ISO beyond the peaks:
The 2nd ISO workshop on analytical spectroscopy'' Eds. A. Salama,
M.F.Kessler, K. Leech \& B. Schulz, ESA-SP 456, p.249 
\reference {} Le Floc'h, E., et al. 2005, ApJ, 632, 169
\reference {} Leger, A., \& Puget, J.-L.  1984, A\&A, 137, L5
\reference {} Li., A. \& Draine, B. T., 2001, ApJ, 554, 778
\reference {} Li., A. \& Draine, B. T., 2002, ApJ, 576, 762
\reference {} Lonsdale, C. J., \& Helou, G. 1987, ApJ, 314, 513
\reference {} Martin, D. C., et al. 2005, ApJ, 619, L1
\reference {} Mortier, A. M. J., et al. 2005, MNRAS, 363, 563
\reference {} Paturel, G., Petit, C., Prugniel, P., Theureau, G., Rousseau,
J., Brouty, M., Dubois, P., \& Cambr{\'e}sy, L.,  2003, A\&A, 412, 45
\reference {} Per\'ez-Gonz\'alez, P., et al. 2008, ApJ, 675, 234
\reference {} Popescu, C. C., \& Tuffs R. J. 2003, A\&A, 410, L21
\reference {}  Press, W. H., Flannery, B. P., Teukolsky, S. A., Vetterling,
W. T. 1986, $Numerical$ $Recipes$, (Cambridge: Cambridge Univ. Press)
\reference {} Puget, J.-L., Abergel, A., Bernard, J.-P., Boulanger,
F., Burton, W. B., D\'esert, F.-X., \& Hartmann, D. 1996, A\&A, 308, 5
\reference {} Reach, W. T. et al. 2005, PASP, 117, 978
\reference {} Rieke, G. H. \& Lebofsky, M. J. 1978, ApJ, 222, 37
\reference {} Rieke, G. H., et al. 2004, ApJS, 154, 25
\reference {} Rieke, G. H., et al. 2008, AJ, 135, 2245
\reference {} Rieke, G. H., et al. 2009, ApJ, accepted, arXiv
0810.4150
\reference {} Sanders, D. B., \& Mirabel, I. F. 1996, ARA\&A, 34, 749
\reference {} Sauvage, M., Thuan, T. X.  1992, ApJ, 396, L69
\reference {} Sauvage, M., Tuffs, R. J., \& Popescu, C. C. 2005,
Sp. Sci. Rev., 119, 313
\reference {} Schuster, M. T., Marengo, M. \* Patten, B. M. 2006,
``Observatory Operations: Strategies, Processes, and Systems''. Eds: 
D. R. Silva \& R. E. Doxsey, Proceedings of the SPIE, Volume 6270, p. 627020
\reference {} Serabyn, E., Lis, D. C., Dowell, C. D., Benford, D. J.,
Hunter, T. R., Trewhella, M., \& Moseley, S. H. 1999, in
``Astrophysics with Infrared Surveys: A Prelude to SIRTF'',
Eds. M.D. Bicay, C. A. Beichman, R. M. Cutri, and B. F. Madore, ASPC,
117, 209
\reference {} Siebenmorgen, R., \& Kr\"ugel, E. 2007, A\&A, 461, 445
\reference {} Smith, J.-D. T., et al. 2007, ApJ, 656, 770
\reference {} Stansberry, J., et al. 2007, PASP, 119, 1038
\reference {} Tuffs, R. J., \& Gabriel, C. 2003, A\&A, 410, 1075
\reference {} Tuffs, R. J., \& Popescu, C. C. 2005, in ``The Spectral
Energy Distribution of Gas-Rich Galaxies: Confronting Models with
Data'', (eds. C.C. Popescu \& R. J. Tuffs), (New York: AIP), p. 344 
\reference {} Vlahakis, C., Dunne, L. \& Eales, S. 2005, MNRAS, 364, 1253
\reference {} Weingartner, \& Draine, B. 2001a, ApJ, 548, 296
\reference {} Weingartner, \& Draine, B. 2001b, ApJS, 134, 263
\reference {} Wen, X.-Q., Wu, H., Cao, C., Xia, X.-Y., 2007, ChJAA, 7, 187 
\end{references}
\end{document}